\documentclass[a4paper,fleqn,useAMS,usenatbib]{mnras}

\usepackage[T1]{fontenc}
\usepackage{ae,aecompl}

\usepackage{amssymb}
\usepackage{aas_macros}
\usepackage{natbib}
\usepackage{times}
\usepackage{graphicx}
\usepackage{xspace}

\newcommand{\sub}[1]{_{\rm #1}}
\newcommand{\up}[1]{^{\rm #1}}

\newcommand{\sfrratio}      {$\mathit{SFR}\sub{rec}/\mathit{SFR}$\xspace}
\newcommand{\mstarratio}    {$M\sub{\ast,rec}/M\sub{\ast}$\xspace}

\title[Recycling of stellar ejecta]{Recycled stellar ejecta as fuel for star formation and implications for the origin of the galaxy mass-metallicity relation}

\author[M. C. Segers et al.]{
Marijke C. Segers,$^{1}$\thanks{E-mail: segers@strw.leidenuniv.nl}
Robert A. Crain,$^{2,1}$
Joop Schaye,$^{1}$
Richard G. Bower,$^{3}$
\newauthor Michelle Furlong,$^{3}$
Matthieu Schaller$^{3}$
and Tom Theuns$^{3}$
\\
$^{1}$Leiden Observatory, Leiden University, PO Box 9513, NL-2300 RA Leiden, the Netherlands\\
$^{2}$Astrophysics Research Institute, Liverpool John Moores University, 146 Brownlow Hill, Liverpool, L3 5RF, UK\\
$^{3}$Institute for Computational Cosmology, Department of Physics, University of Durham, South Road, Durham DH1 3LE, UK
}

\date{Accepted 2015 October 29. Received 2015 October 28; in original form 2015 October 28}

\pubyear{2015}

\begin{document}
\label{firstpage}
\pagerange{\pageref{firstpage}--\pageref{lastpage}}
\maketitle

\begin{abstract}
We use cosmological, hydrodynamical simulations from the EAGLE and OWLS projects to assess the significance of recycled stellar ejecta as fuel for star formation. The fractional contributions of stellar mass loss to the cosmic star formation rate (SFR) and stellar mass densities increase with time, reaching $35 \%$ and $19 \%$, respectively, at $z=0$. The importance of recycling increases steeply with galaxy stellar mass for $M\sub{\ast} < 10^{10.5}$ M$\sub{\odot}$, and decreases mildly at higher mass. This trend arises from the mass dependence of feedback associated with star formation and AGN, which preferentially suppresses star formation fuelled by recycling. Recycling is more important for satellites than centrals and its contribution decreases with galactocentric radius. The relative contribution of AGB stars increases with time and towards galaxy centers. This is a consequence of the more gradual release of AGB ejecta compared to that of massive stars, and the preferential removal of the latter by star formation-driven outflows and by lock up in stellar remnants. Recycling-fuelled star formation exhibits a tight, positive correlation with galaxy metallicity, with a secondary dependence on the relative abundance of alpha elements (which are predominantly synthesized in massive stars), that is insensitive to the subgrid models for feedback. Hence, our conclusions are directly relevant for the origin of the mass-metallicity relation and metallicity gradients. Applying the relation between recycling and metallicity to the observed mass-metallicity relation yields our best estimate of the mass-dependent contribution of recycling. For centrals with a mass similar to that of the Milky Way, we infer the contributions of recycled stellar ejecta to the SFR and stellar mass to be $35 \%$ and $20 \%$, respectively.
\end{abstract}



\section{Introduction}
\label{sec:introduction}

The rate at which galaxies form stars is closely related to the amount of fuel that is available. Although we still lack a complete understanding of how galaxies obtain their gas, several potential sources of star formation fuel have been investigated in previous works, both observationally and using hydrodynamical simulations \citep[e.g.][]{putman_2009}. Galaxies accrete gas from the intergalactic medium (IGM) along cold, dense, filamentary streams \citep[e.g.][]{keres_2005,dekel_2009,brooks_2009,vandevoort_2012}, which can extend far inside the halo virial radius, and through quasi-spherical infall from a diffuse hot halo, which contains gas that has been shock-heated to the halo virial temperature \citep{rees_ostriker_1977,silk_1977}. Cosmological, hydrodynamical simulations give predictions for the relative importance of these two `modes' of gas accretion, generally indicating a dominant role for the cold mode in the global build up of galaxies, with the hot mode becoming increasingly important towards lower redshifts and in more massive systems \citep[e.g.][]{birnboim+dekel_2003,keres_2005,keres_2009,crain_2010,vandevoort_2011,nelson_2013}. Galaxies can also acquire new fuel for star formation by stripping the gas-rich envelopes of merging galaxies as soon as these become satellites in a group or cluster environment \citep[e.g.][]{sancisi_2008,vandevoort_2011} or by re-accreting gas that has previously been ejected from the galaxy in an outflow and is raining back down in the form of a halo fountain \citep[e.g.][]{oppenheimer+dave_2008,oppenheimer+dave_2010}.

In addition to the various channels of accreting gas from the IGM, every galaxy has an internal channel for replenishing the reservoir of gas in the interstellar medium (ISM), namely the shedding of mass by the stellar populations themselves. Stars lose a fraction of their mass in stellar winds before and while they go through the asymptotic giant branch (AGB) phase. Furthermore, a substantial amount of stellar material is released as stars end their lives in supernova (SN) explosions. Eventually, $\sim 50 \%$ of the initial mass of a stellar population will be released. If this material is not ejected into the circumgalactic medium (CGM), where it can emerge as X-ray emitting gas in the hot circumgalactic corona \citep[e.g.][]{parriott+bregman_2008,crain_2013}, or entirely expelled from the galaxy into the IGM \citep[e.g.][]{ciotti_1991}, but rather ends up in the cool ISM gas reservoir, then it can be `recycled' to fuel subsequent generations of star formation \citep[e.g.][]{mathews_1990,martin_2007}. Note that what we call `gas recycling' here is different from the process considered in works on galactic outflow fountains, in which `recycling' refers to the re-accretion of gas ejected from the ISM, regardless of whether it has ever been part of a star. In this work `recycled gas' refers to the gas from evolved stars that is used to form new stars, regardless of whether it has been blown out of a galaxy.

Using observational constraints on the rates of gas infall and the history of star formation, \citet{leitner+kravtsov_2011} assessed the significance of recycled stellar evolution products in the gas budget of a number of nearby disk galaxies (including the Milky Way). They modeled the global mass loss history of each galaxy from an empirically motivated distribution of stellar population ages and a set of stellar yields and lifetimes, and showed that the gas from stellar mass loss can provide most of the fuel required to sustain the current rates of star formation. They suggested that this internal supply of gas is important for fuelling star formation at late epochs, when the cosmological accretion rate drops or is suppressed by preventative feedback \citep[e.g.][]{mo+mao_2002}, hence falling short of the observed star formation rate (SFR) of the galaxies. Furthermore, \citet{voit+donahue_2011} argued that due to the high ambient pressures and the resulting short gas cooling times, central cluster galaxies are very efficient at recycling stellar ejecta into new stars. They showed that the stellar mass loss rates are generally comparable to, or even higher than, the observed rates of star formation and emphasized the importance of including this form of internal gas supply in any assessment of the gas budget of such systems. These conclusions are consistent with the observation by \citet{kennicutt_1994} that recycling of stellar ejecta can extend the lifetimes of gaseous discs by factors of $1.5 - 4$, enabling them to sustain their ongoing SFRs for periods comparable to the Hubble time \citep[see also][]{roberts_1963,sandage_1986}. These studies suggest that recycled stellar mass loss is an important part of the gas budget of star-forming galaxies, even hinting that it may be a necessary ingredient to reconcile the gas inflow and consumption rates of the Milky Way.

In this paper, we investigate the importance of gas recycling for fuelling star formation by explicitly calculating the contribution of stellar mass loss to the SFR and stellar mass of present-day galaxies. We use cosmological simulations from the Evolution and Assembly of GaLaxies and their Environments (EAGLE) project (\citealt{schaye_2015}, hereafter S15; \citealt{crain_2015}) to explore the recycling of stellar ejecta, as a cosmic average as a function of redshift and within individual (central and satellite) galaxies at $z=0$, where we give quantitative predictions for recycling-fuelled star formation as a function of galaxy stellar mass and establish a connection with observational diagnostics by relating these predictions to gas-phase and stellar metallicities.

The EAGLE simulations explicitly follow the mass released by stellar populations in the form of stellar winds and SN explosions of Types Ia and II, enabling us to study the relative significance of these mass loss channels for fuelling star formation. The subgrid parameters in the models for feedback associated with star formation and active galactic nuclei (AGN) have been calibrated to reproduce the $z \simeq 0$ observed galaxy stellar mass function (GSMF) and the relation between stellar mass, $M\sub{\ast}$, and the mass of the central supermassive black hole (BH), $M\sub{BH}$, with the additional constraint that the sizes of disc galaxies must be reasonable. The EAGLE simulations not only successfully reproduce these key observational diagnostics with unprecedented accuracy, but are also in good agreement with a large and representative set of low- and high-redshift observables that were not considered during the calibration \citep[S15,][]{crain_2015,furlong_2015,lagos_2015,rahmati_2015,sawala_2015,schaller_2015a,trayford_2015}.

We consider the reproduction of a realistic galaxy population to be a prerequisite for this study, since its conclusions are sensitive to the detailed evolution of the gas `participating' in galaxy formation, requiring that the simulations accurately model the evolving balance between the inflow of gas onto galaxies and the combined sinks of star formation and ejective feedback. That EAGLE satisfies this criterion is particularly advantageous, since hydrodynamical simulations are not subject to several limiting approximations inherent to simpler techniques, for example semi-analytic models of galaxy formation. This, in addition to their inclusion of a detailed implementation of chemodynamics, makes the EAGLE simulations an ideal tool for establishing quantitative predictions concerning the role of gas recycling in fuelling star formation.

We also briefly explore the sensitivity of our results to the physical processes in the subgrid model. To do so, we use a suite of cosmological simulations from the OverWhelmingly Large Simulations (OWLS) project \citep{schaye_2010}. As the OWLS project aimed to explore the role of the different physical processes modelled in the simulations, it covers a wide range of subgrid implementations and parameter values, including extreme variations of the feedback model and variations of the stellar initial mass function (IMF). We will show that the efficiency of the feedback associated with star formation and AGN plays an important role in regulating the fuelling of star formation with recycled stellar ejecta.

We note that, because of the tight correlation we find between the contribution of stellar mass loss to the SFR (stellar mass) and the ISM (stellar) metallicity, our characterization and explanation of the role of stellar mass loss as a function of galaxy mass and type has important and direct implications for the origin of the mass-metallicity relation.

This paper is organized as follows. In Section~\ref{sec:simulations} we present a brief overview of the simulation set-up and the subgrid modules implemented in EAGLE. In this section we also introduce the two quantities we use to assess the importance of gas recycling, namely the fractional contributions of stellar mass loss to the SFR and stellar mass. In Section~\ref{sec:eagle} we present quantitative predictions from EAGLE for the evolution of the cosmic averages of these quantities and for their dependence on metallicity and galaxy stellar mass. We explore the sensitivity of these results using a set of OWLS simulations in Section~\ref{sec:massdep_owls}. Finally, we summarize our findings in Section~\ref{sec:summary}.


\section{Simulations}
\label{sec:simulations}

The amount of gas that galaxies can recycle to form new generations of stars, depends fundamentally on the fraction of stellar mass that is returned to the ISM. How much of this mass is actually used to fuel star formation is not straightforward to calculate analytically, due to the variety of processes, such as cosmological infall, gas stripping of satellite galaxies, and feedback associated with star formation and AGN, that can have an effect on the star formation histories of individual galaxies. Hence, we use cosmological simulations from the EAGLE and OWLS projects to investigate this.

For the majority of this work we use the EAGLE simulations, which were run with a modified version of the smoothed particle hydrodynamics (SPH) code \textsc{Gadget3} \citep[last described by][]{springel_2005} using a pressure-entropy formulation of SPH (\citealt{hopkins_2013}; see \citealt{schaller_2015b} for a comparison between SPH flavours). The simulations adopt a $\Lambda$CDM cosmology with parameters $\left[ \Omega\sub{m},\Omega\sub{b},\Omega\sub{\Lambda},\sigma\sub{8},n\sub{s},h \right]=\left[ 0.307,0.04825,0.693,0.8288,0.9611,0.6777 \right]$ \citep{planck_2014}.

We will study primarily the largest EAGLE simulation, which we will refer to as \emph{Ref-L100N1504} (as in S15) or as the `fiducial' model. This simulation was run in a periodic volume of size $L = 100$ comoving Mpc (cMpc), containing $N = 1504^3$ dark matter particles and an equal number of baryonic particles. The gravitational softening length of these particles is $2.66$ comoving kpc (ckpc), limited to a maximum physical scale of $0.7$ proper kpc (pkpc). The particle masses for baryons and dark matter are initially $m\sub{b}=1.8 \times 10^6\ {\rm M}\sub{\odot}$ and $m\sub{dm}=9.7 \times 10^6\ {\rm M}\sub{\odot}$, respectively. However, during the course of the simulation the baryonic particle masses change as mass is transferred from star to gas particles, corresponding to the recycling of mass from stellar populations back into the gas reservoir.


\subsection{Subgrid physics}
\label{sec:subgrid_physics}

The subgrid physics used in EAGLE is largely based on the set of subgrid recipes developed for OWLS, but includes a few important improvements. Star formation is modelled using a metallicity-dependent density threshold \citep[given by][]{schaye_2004}, above which gas particles are assigned a pressure-dependent SFR \citep[that by construction reproduces the observed Kennicutt-Schmidt star formation law;][]{schaye+dallavecchia_2008} and are converted stochastically into star particles. Each star particle represents a stellar population of a single age (simple stellar population; SSP) and inherits its mass and metallicity from its progenitor gas particle. The adopted IMF is a \citet{chabrier_2003} IMF, spanning the mass range of $0.1-100$ M$\sub{\odot}$. Following the prescriptions of \citet{wiersma_2009b}, an SSP loses mass through stellar winds and supernova explosions (SN Type II) from massive stars and through AGB winds and SN Type Ia explosions from intermediate-mass stars. The time-dependent mass loss, which we show in Section~\ref{sec:ssp}, is calculated using the metallicity-dependent stellar lifetime tables of \citet{portinari+chiosi+bressan_1998}, in combination with the set of nucleosynthetic yields of \citet{marigo_2001} (for stars in the mass range $0.8-6$ M$\sub{\odot}$ ) and \citet{portinari+chiosi+bressan_1998} (for stars in the mass range $6-100$ M$\sub{\odot}$), all of which are based on the same Padova evolutionary tracks. For SN Type Ia, the yields are taken from the W7 model of \citet{thielemann_2003} and the distribution of progenitor lifetimes is modelled using an empirically motivated time-delay function that is calibrated to reproduce the observed cosmic SN Type Ia rate (see fig. 3 of S15). At every gravitational time step (every $10$th time step for star particles older than $100$ Myr), the ejecta are distributed over the neighbouring gas particles according to the SPH interpolation scheme\footnote{As discussed in S15 and different from \citet{wiersma_2009b}, EAGLE uses weights that are independent of the current gas particle mass for the distribution of stellar mass loss. The reason for this is to avoid a runaway process, causing a small fraction of the particles to end up with very large masses compared to their neighbours, as particles that have grown massive due to enrichment, are also likely to become increasingly enriched in future time steps, if they carry more weight in the interpolation.}. The simulations follow the abundances of $11$ individual elements, which are used to calculate the rates of radiative cooling and heating on an element-by-element basis and in the presence of \citet{haardt+madau_2001} UV and X-ray background radiation \citep{wiersma_2009a}. Energy feedback from star formation is implemented by stochastically injecting thermal energy into the gas surrounding newly-formed star particles as described by \citet{dallavecchia+schaye_2012}. The fraction $f\sub{th}$ of the total available feedback energy that is used to heat the gas, depends on the local gas metallicity and density, so as to account for increased thermal losses in higher metallicity gas and to compensate for the increased numerical radiative losses in higher density gas \citep{crain_2015}. The growth of BHs is modelled by inserting seed BHs into haloes more massive than $m\sub{halo,min}=10^{10}\ h^{-1}\ {\rm M}\sub{\odot}$, which can grow either through gas accretion, at a rate that depends on the angular momentum of the gas, or through mergers with other BHs \citep{booth+schaye_2009,rosasguevara_2015}. AGN feedback is implemented as the stochastic injection of thermal energy into the gas surrounding the BH \citep{booth+schaye_2009,dallavecchia+schaye_2012}. The subgrid routines for stellar and AGN feedback have been calibrated to reproduce observations of the present-day GSMF, the $M\sub{\ast} - M\sub{BH}$ relation and to yield reasonable galaxy sizes \citep[S15,][]{crain_2015}.


\subsection{Mass released by an SSP}
\label{sec:ssp}

\begin{figure*}
\begin{center}
\includegraphics[width=0.95\textwidth]{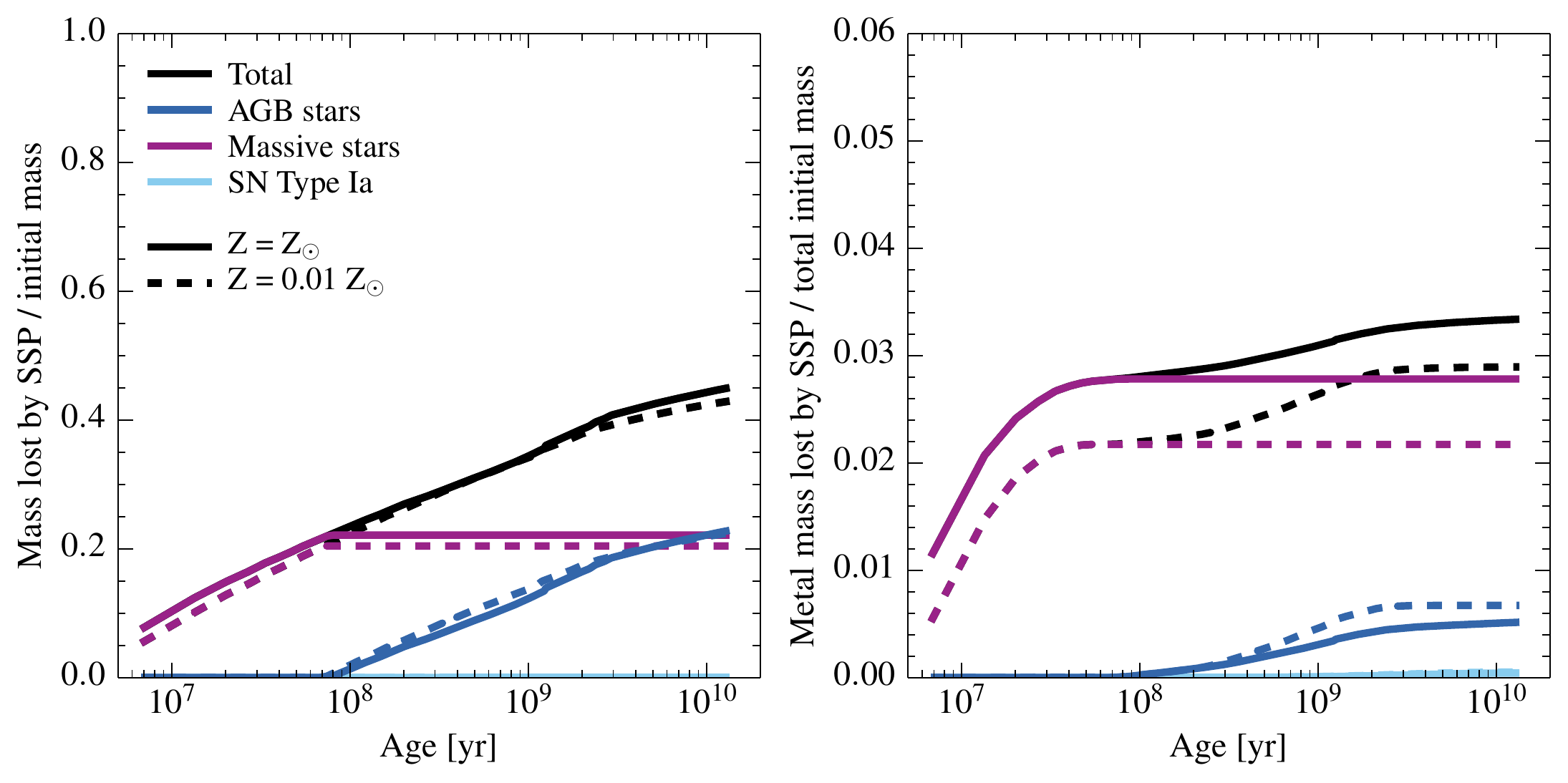}
\end{center}
\caption{The cumulative fraction of the initial mass (total: left panel; in the form of metals, i.e. elements heavier than helium: right panel) that is released by an SSP as a function of its age, adopting a \citet{chabrier_2003} IMF in the range $0.1-100$ M$\sub{\odot}$. The curves show the contributions from AGB stars (blue), massive stars (purple) and SN Type Ia (cyan), as well as the total (metal) mass ejected by the SSP (black), for two stellar metallicities: solar (solid) and $1$ percent of solar (dashed). Initially, only massive stars contribute to the mass loss, but for SSP ages $\gtrsim 10^{8}$ yr the contribution from AGB stars becomes increasingly significant. These AGB ejecta are, however, less metal-rich than the ejecta from massive stars. The contribution from SN Type Ia to the (metal) mass loss remains insignificant at all times. Increasing the metallicity does not have a strong effect on the total mass loss, but increases the total ejected metal mass as well as the relative contribution from massive stars.}
\label{fig:SSP_chabrier}
\end{figure*}

Fig.~\ref{fig:SSP_chabrier} shows the total mass (left panel) and metal mass (right panel) released by an SSP as a function of its age as prescribed by the chemodynamics model. The curves show the total integrated mass ejected (black) and the integrated mass sourced by AGB stars (blue), massive stars (i.e. stellar winds plus SN Type II; purple) and SN Type Ia (cyan) for two different SSP metallicities: solar (solid lines) and $1$ percent of solar (dashed), using a solar value of Z$\sub{\odot} = 0.0127$. Both panels show that the ejected (metal) mass, which is expressed as a fraction of the total initial mass of the SSP, increases as the SSP ages. Initially only massive stars contribute, but for ages $\gtrsim 10^8$ yr the contribution from AGB stars becomes increasingly significant. Comparing, for each channel, the total to the metal mass loss shows that the ejecta from massive stars are more metal-rich than those from AGB stars. The contribution from SN Type Ia to the (metal) mass loss remains insignificant for all SSP ages\footnote{Note that we only show the relative contributions from massive and intermediate-mass stars to the \emph{total} ejected metal mass. These may be different from their contributions to the ejected mass of individual elements, as for example iron, which has a substantial fraction of its abundance sourced by SN Type Ia \citep[see fig.2 of][]{wiersma_2009b}.}. Varying the metallicity over two orders of magnitude changes the total mass loss by only a few percent, but changes the total ejected metal mass, as well as the relative contribution from massive stars, by $\sim 10 - 15 \%$.

Since the choice of IMF determines the relative mass in intermediate-mass and massive stars per unit stellar mass formed, it affects the mass loss from an SSP. \citet{leitner+kravtsov_2011} indeed show that the differences between alternative, reasonable choices of the IMF can be significant (see their fig. 1). In Appendix~\ref{sec:ssp_salpeter}, we similarly conclude that the total and metal mass loss is a factor of $\sim 1.5$ greater for a Chabrier IMF than for the more bottom-heavy Salpeter IMF (which is adopted by one of the OWLS model variations examined in Section~\ref{sec:massdep_owls}).


\subsection{Numerical convergence}
\label{sec:convergence}

In order to test for numerical convergence, we use a set of three simulations that were run in volumes of size $L = 25$ cMpc. This includes a high-resolution simulation (\emph{Recal-L025N0752}), whose subgrid feedback parameters were recalibrated to improve the fit to the observed present-day GSMF (see table 3 of S15). We show a concise comparison between the fiducial simulation and \emph{Recal-L025N0752} when we present results as a function of halo and stellar mass in Section~\ref{sec:massdep_eagle}, while a more detailed convergence test can be found in Appendix~\ref{sec:resotests_massdep}. In the rest of the results section (Section~\ref{sec:eagle}) we use only the fiducial simulation, which, due to its $64$ times greater volume than \emph{Recal-L025N0752}, provides a better statistical sample of the massive galaxy population, and models a more representative cosmic volume.


\subsection{Identifying haloes and galaxies}
\label{sec:halo_galaxy}

Haloes are identified using a Friends-of-Friends (FoF) algorithm \citep{davis_1985}, linking dark matter particles that are separated by less than $0.2$ times the mean interparticle separation. Gas and star particles are assigned to the same halo group as their nearest dark matter particle. The \textsc{subfind} algorithm \citep{springel_2001,dolag_2009} then searches for gravitationally bound substructures within the FoF haloes, which we label `galaxies' if they contain stars. The galaxy position is defined to be the location of the particle with the minimum gravitational potential within the subhalo. The galaxy at the absolute minimum potential in the FoF halo (which is almost always the most massive galaxy) is classified as the `central' galaxy, whereas the remaining subhaloes are classified as `satellite' galaxies.

The mass of the main halo, $M\sub{200}$, is defined as the mass internal to a spherical shell centred on the minimum gravitational potential, within which the mean density equals $200$ times the critical density of the Universe. The subhalo mass, $M\sub{sub}$, corresponds to all the mass bound to the substructure identified by \textsc{subfind}. The stellar mass, $M\sub{\ast}$, refers to the total mass in stars that is bound to this substructure and that resides within a 3D spherical aperture of radius $30$ pkpc. Other galaxy properties, such as the SFR and the stellar half-mass radius, are also computed considering only particles within this aperture, mimicking observational measurements of these quantities (as shown in fig 6. of S15, the present-day GSMF using a $30$ pkpc 3D aperture is nearly identical to the one using the 2D Petrosian aperture applied by SDSS). The aperture has negligible effects on stellar masses for $M\sub{\ast} < 10^{11}$ M$\sub{\odot}$ and galactic SFRs, as the vast majority of the star formation takes place within the central 30 pkpc. For the more massive galaxies, on the other hand, the stellar masses are somewhat reduced, as the aperture cuts out the diffuse stellar mass at large radii that would contribute to the intracluster light.


\subsection{Measuring the star formation rate and stellar mass contributed by recycling}
\label{sec:recycling}

We explicitly track the contributions to the SFR and stellar mass from gas recycling. For a gas particle of mass $m\sub{g}^i(t)$ at time $t$, the total fraction of its mass contributed by released stellar material (in the form of hydrogen, helium and heavy elements) is given by
\begin{equation}
f\sub{g,rec}^i(t)=\frac{m\sub{g}^i(t)-m\sub{b}}{m\sub{g}^i(t)},
\end{equation}
where $m\sub{b}$ is the initial gas mass of gas particles at the start of the simulation. Since a gas particle is the smallest quantum of mass we are able to consider, its recycled fraction is by construction assumed to be perfectly mixed. Therefore, if the gas particle is considered star-forming, $f\sub{g,rec}^i(t)$ also indicates the fraction of its current SFR that is contributed by stellar ejecta. Then, summing up the contributions from all $N\sub{g}\up{gal}$ gas particles in a galaxy (within the $30$ pkpc 3D aperture) yields the SFR contributed by recycling for this galaxy:
\begin{equation}\label{eq:SFRfromgas_gal}
\mathit{SFR}\sub{rec}\up{gal}(t)=\sum\sub{i=1}^{N\sub{g}\up{gal}}\frac{m\sub{g}^i(t)-m\sub{b}}{m\sub{g}^i(t)}\mathit{SFR}^i(t),
\end{equation}
where $\mathit{SFR}^i(t)$ is the SFR of gas particle $i$ at time $t$. Similarly, summing up the contributions from all $N\sub{g}\up{cos}$ gas particles in the simulation volume yields the cosmic average of this quantity:
\begin{equation}\label{eq:SFRfromgas_cos}
\mathit{SFR}\sub{rec}\up{cos}(t)=\sum\sub{i=1}^{N\sub{g}\up{cos}}\frac{m\sub{g}^i(t)-m\sub{b}}{m\sub{g}^i(t)}\mathit{SFR}^i(t).
\end{equation}

Since a star particle inherits its mass and elemental abundances from its progenitor gas particle, the fraction of its mass contributed by recycling is:
\begin{equation}\label{eq:rec_frac_star}
f\sub{\ast,rec}^j(t)=\frac{m\sub{\ast,init}^j-m\sub{b}}{m\sub{\ast,init}^j},
\end{equation}
where $m\sub{\ast,init}^j = m\sub{g}^j(t\sub{birth})$ is the mass of star particle $j$ at the time of its birth, $t\sub{birth}$. Note that equation~(\ref{eq:rec_frac_star}) is valid for all $t \geq t\sub{birth}$, even though the star particle itself loses mass. This is again a consequence of the assumption of perfect mixing on the particle scale. Summing up the contributions from all $N\sub{\ast}\up{gal}$ star particles in a galaxy that are within the 3D aperture,
\begin{equation}\label{eq:Mstar_gal}
M\sub{\ast,rec}\up{gal}(t)=\sum\sub{j=1}^{N\sub{\ast}\up{gal}}\frac{m\sub{\ast,init}^j-m\sub{b}}{m\sub{\ast,init}^j}m\sub{\ast}^j(t),
\end{equation}
and all $N\sub{\ast}\up{cos}$ star particles in the simulation volume,
\begin{equation}\label{eq:Mstar_cos}
M\sub{\ast,rec}\up{cos}(t)=\sum\sub{j=1}^{N\sub{\ast}\up{cos}}\frac{m\sub{\ast,init}^j-m\sub{b}}{m\sub{\ast,init}^j}m\sub{\ast}^j(t),
\end{equation}
give the galaxy stellar mass and cosmic stellar mass, respectively, contributed by recycled gas.

While $\mathit{SFR}\sub{rec}$ and $M\sub{\ast,rec}$ are related, it is still helpful to consider both: $\mathit{SFR}\sub{rec}$ indicates the instantaneous impact of gas recycling, whereas $M\sub{\ast,rec}$ indicates the importance of recycling over the past history of star formation. In this work we mainly focus on the \emph{relative} contribution of gas recycled from stellar mass loss to the total (cosmic or galactic) SFR and stellar mass. Normalizing $\mathit{SFR}\sub{rec}$ and $M\sub{\ast,rec}$ by the respective total quantities, yields \sfrratio and \mstarratio, specifying the \emph{fractions} of the SFR and the stellar mass that are due to stellar mass loss.

In addition to the total amount of recycling, we will also consider the relative contributions from the different sources of stellar mass loss that were included in the subgrid model (Section~\ref{sec:ssp}). As the transfer of mass from AGB stars, SN Type Ia and massive stars between star and gas particles is explicitly followed by the EAGLE simulations\footnote{Note that these enrichment channels only refer to the \emph{last} enrichment episode. Every stellar population releases mass via the different channels in a way that depends only on its age and metallicity (for a given IMF).}, we can calculate \sfrratio and \mstarratio solely due to gas from AGB stars by simply replacing $m\sub{g}^i(t)-m\sub{b}$ in equations~(\ref{eq:SFRfromgas_gal}) and (\ref{eq:SFRfromgas_cos}) by $m\sub{AGB}^i$ and replacing $m\sub{\ast,init}^j-m\sub{b}$ in equations~(\ref{eq:Mstar_gal}) and (\ref{eq:Mstar_cos}) by $m\sub{AGB,init}^j$, where $m\sub{AGB}$ is the mass from AGB stars in the respective gas or star particle. The \sfrratio and \mstarratio due to gas from SN Type Ia and massive stars are calculated analogously.


\section{Recycled stellar mass loss in EAGLE}
\label{sec:eagle}

In this section we use the fiducial EAGLE simulation, \emph{Ref-L100N1504}, to make quantitative predictions for the importance of gas recycling for fuelling ongoing star formation in present-day galaxies over a wide range of galaxy masses. However, we start with a brief investigation of the evolution of recycling-fuelled star formation over cosmic history.


\subsection{Evolution of the cosmic average}
\label{sec:evolution}

\begin{figure*}
\begin{center}
\includegraphics[width=0.95\textwidth]{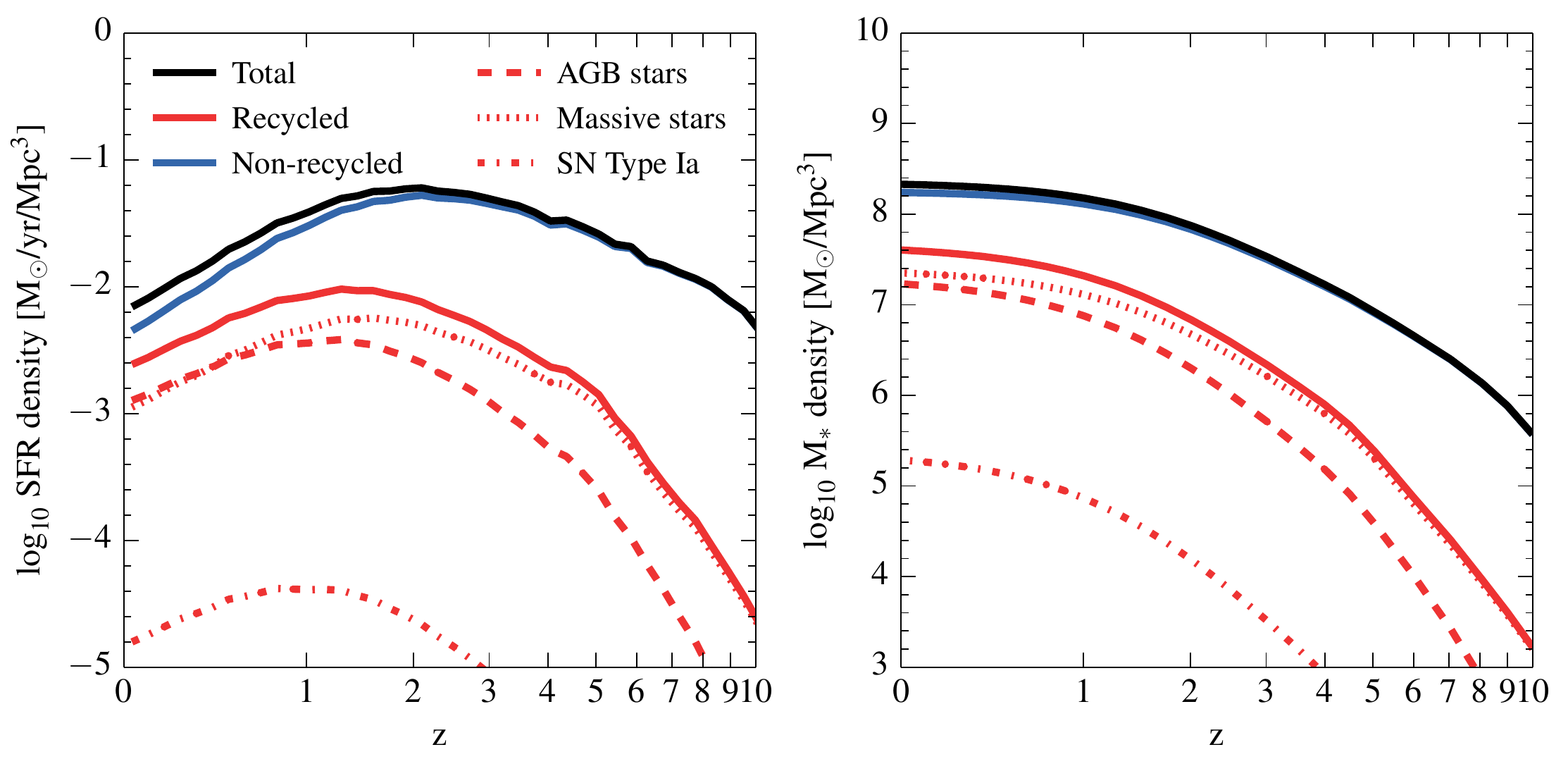}
\end{center}
\caption{The evolution of the cosmic SFR density (left) and the cosmic stellar mass density (right) fuelled by recycled stellar mass loss (red), as well as the SFR and stellar mass densities fuelled by all gas (black) and gas that has not been recycled (blue). The `recycled' SFR and stellar mass densities are split according to the contributions from AGB stars (dashed), massive stars (dotted) and SN Type Ia (dot-dashed). Recycling of stellar mass loss becomes increasingly important for fuelling star formation towards the present day. The gas from massive stars accounts for the majority of the cosmic SFR and stellar mass density from recycled gas at high redshift, but the contribution from AGB stars increases with time (accounting for the majority of the `recycled' SFR density for $z \lesssim 0.4$).}
\label{fig:evo_sfr_mstar}
\end{figure*}

\begin{figure}
\begin{center}
\includegraphics[width=\columnwidth]{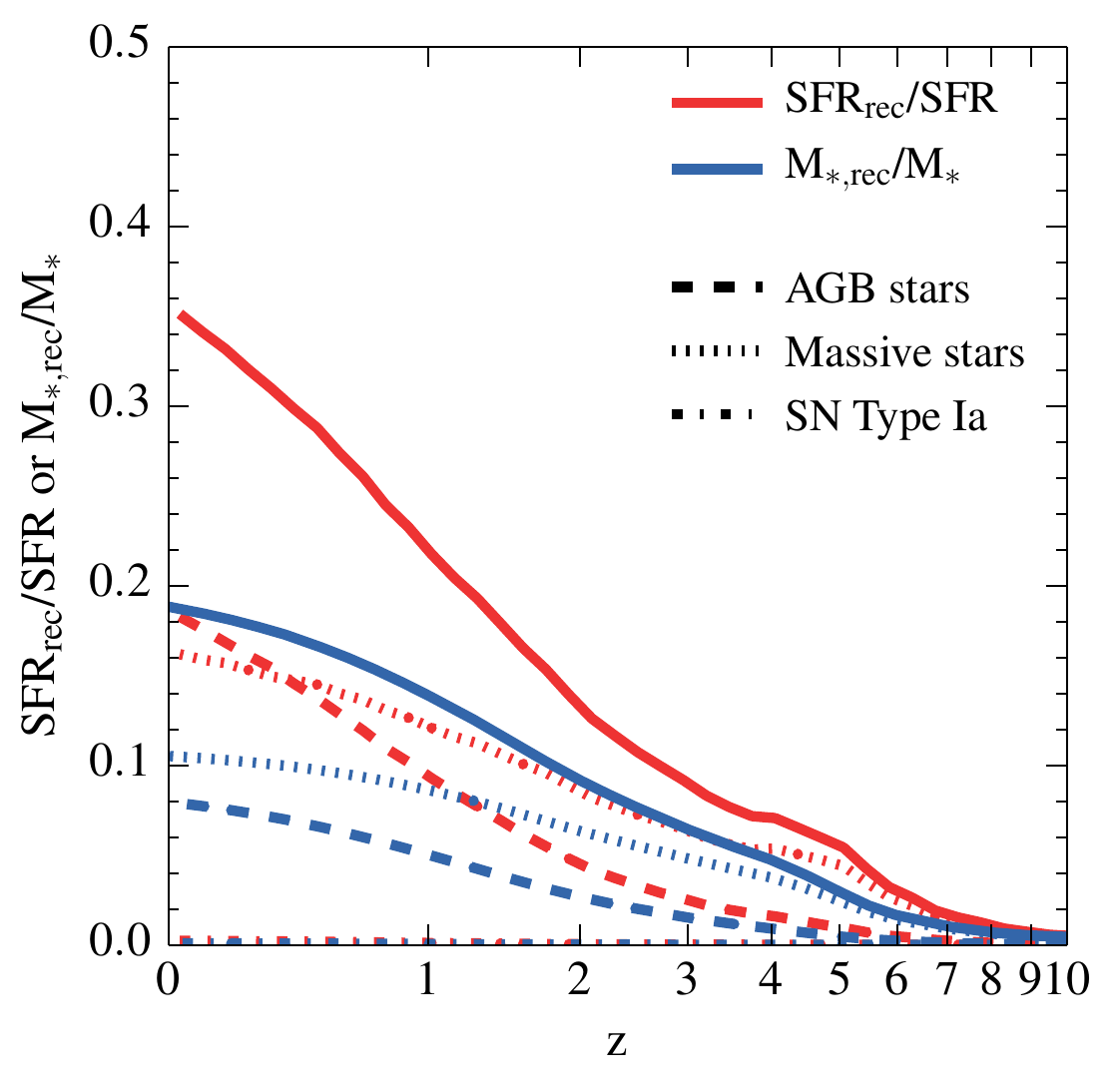}
\end{center}
\caption{The evolution of the fractional contribution of recycled stellar mass loss to the cosmic SFR density (red) and cosmic stellar mass density (blue), where we show the total (solid) as well as the contributions from AGB stars (dashed), massive stars (dotted) and SN Type Ia (dot-dashed). With decreasing redshift, an increasing fraction of the cosmic SFR and stellar mass density is fuelled by recycled gas, which we find to be $35 \%$ and $19 \%$, respectively, at $z = 0$.}
\label{fig:evo_ratios}
\end{figure}

The left panel of Fig.~\ref{fig:evo_sfr_mstar} shows the total cosmic SFR density (black), the cosmic SFR density fuelled by stellar mass loss (red, solid: `recycled') and the cosmic SFR density fuelled by unprocessed gas (blue: `non-recycled') as a function of redshift. The red curve has been split into the contributions from the three mass loss channels that are tracked by the simulation: AGB stars (dashed), massive stars (dotted) and SN Type Ia (dot-dashed). To get a better idea of the evolution of the \emph{fractional} contribution from recycled gas to the cosmic star formation history, we show the evolution of the cosmic average \sfrratio, as well as the fractional contribution per channel, in Fig.~\ref{fig:evo_ratios} (red).

At $z > 2$ there is little difference between the SFR density due to `non-recycled' gas and the total SFR density. At these high redshifts most of the fuel for star formation is due to unprocessed gas\footnote{Note that this does not imply that most of the SFR, and hence stellar mass, is in the form of Pop III (i.e. metal-free) stars, because the stellar evolution products are mixed with the unprocessed material (in the simulations on the scale of a gas particle).}, since there has simply not been much time for stellar populations to evolve and to distribute a significant amount of gas that can be recycled. From the `recycled' curve we see that the SFR density fuelled by recycled stellar mass loss rises rapidly at high redshift, peaks at $z \approx 1.3$, and then declines steadily towards $z = 0$. This trend is similar to the evolution of the total SFR density, although with a delay of $\sim 1.5$ Gyr (the total SFR density peaks at $z \approx 2$). Furthermore, the slope of the `recycled' curve is steeper at high redshift and shallower at low redshift compared to that of the total SFR density, indicating that gas recycling becomes increasingly important for fuelling star formation. This is consistent with the rapid rise of the total \sfrratio with decreasing redshift in Fig.~\ref{fig:evo_ratios}. Our fiducial EAGLE model indicates that $35 \%$ of the present-day cosmic SFR density is fuelled by recycled stellar mass loss.

The right panel of Fig.~\ref{fig:evo_sfr_mstar} shows the build up of the cosmic stellar mass density, the total as well as the contributions from recycled and unprocessed gas. The evolution of the cosmic average \mstarratio is shown in Fig.~\ref{fig:evo_ratios} (blue). The stellar mass density is related to the SFR density, as one can calculate the former by integrating the latter over time (while taking into account stellar mass loss). Hence, similar to the SFR density, the stellar mass density is initially ($z \gtrsim 2$) dominated by star formation from unprocessed gas, while the contribution from recycling becomes increasingly important towards $z=0$. EAGLE indicates that, at the present day, $19 \%$ of the cosmic stellar mass density has been formed from recycled stellar mass loss.

Comparing the different sources of stellar mass loss, we see that massive stars initially account for the majority of the SFR and stellar mass density from recycled gas. These stars have short lifetimes and are therefore the first to contribute to the mass loss from a stellar population (see Fig.~\ref{fig:SSP_chabrier}). Towards lower redshift the mass lost by AGB stars becomes increasingly important and even becomes the dominant contributor to the SFR density from recycled gas for $z \lesssim 0.4$ (while remaining subdominant in the case of the stellar mass density). As expected from Fig.~\ref{fig:SSP_chabrier}, recycled SN Type Ia ejecta do not contribute significantly to the cosmic SFR density at any redshift.

\subsection{Relation with metallicity at $z=0$}
\label{sec:metallicity}

\begin{figure*}
\begin{center}
\includegraphics[width=0.95\textwidth]{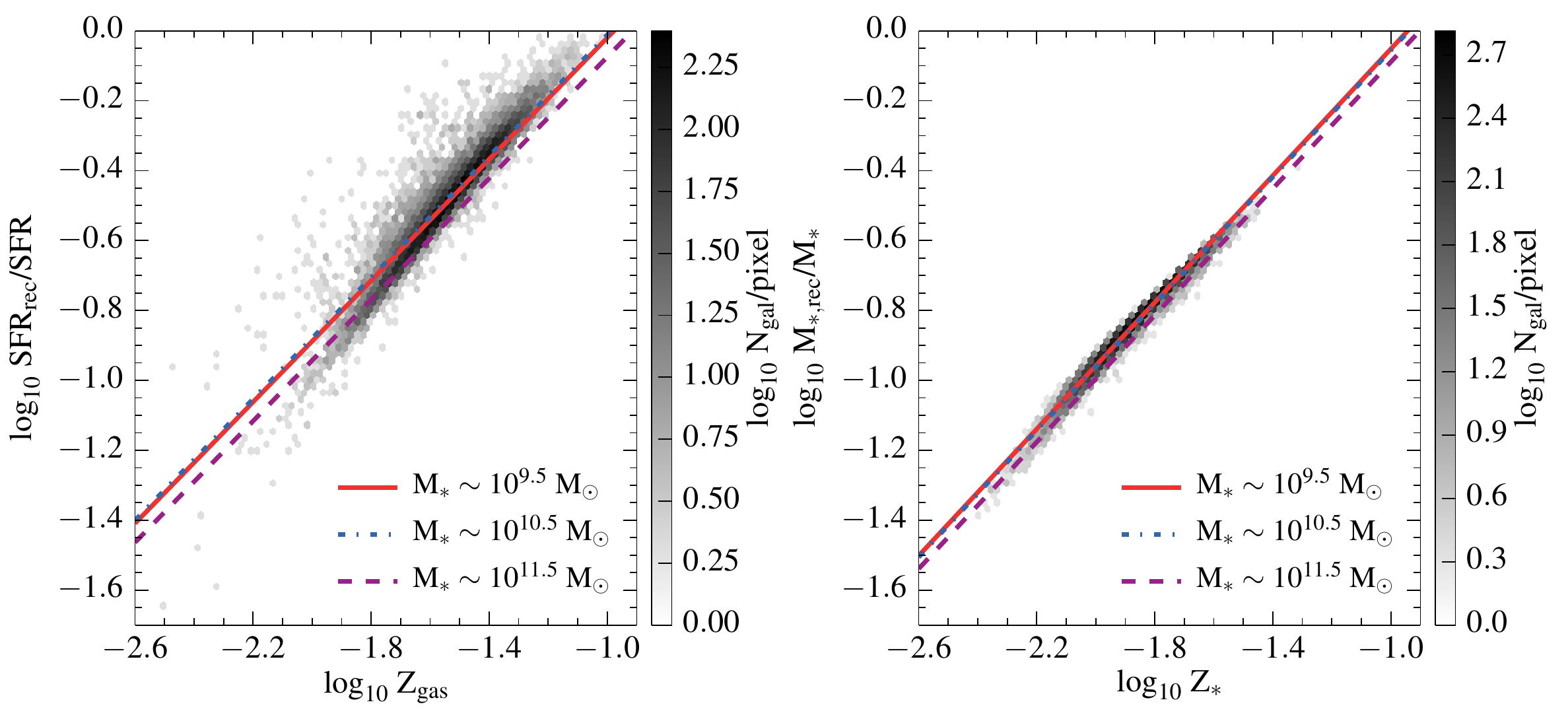}
\end{center}
\caption{The fractional contribution of recycled stellar mass loss to the SFR (left) and stellar mass (right) of central galaxies at $z=0$ as a function of their average ISM and stellar metallicity, respectively. The grey scale indicates the number of galaxies in each cell, where we only include galaxies with stellar masses corresponding to at least $100$ gas particles. In the left panel we only consider subhaloes with a non-zero SFR. We find tight power-law relations between the recycled gas contributions and the respective metallicity measures. These relations exhibit a slight mass dependence as a result of the increasing contribution from massive stars relative to intermediate-mass stars to the SFR and stellar mass for $M\sub{\ast} \gtrsim 10^{10.5}$ M$\sub{\odot}$. The best-fit relations (equations~\ref{eq:sfr_metal} and \ref{eq:mstar_metal}), plotted for galaxies with $M\sub{\ast} \sim 10^{9.5}$ M$\sub{\odot}$ (red, solid line), $M\sub{\ast} \sim 10^{10.5}$ M$\sub{\odot}$ (blue, dot-dashed line) and $M\sub{\ast} \sim 10^{11.5}$ M$\sub{\odot}$ (purple, dashed line), enable one to estimate the importance of gas recycling in present-day galaxies from their observed metallicity and $\alpha$-enhancement.}
\label{fig:metallicity}
\end{figure*}

Having studied the evolution of the cosmic average \sfrratio and \mstarratio, we will now take a closer look at the $z = 0$ values for individual galaxies in the \emph{Ref-L100N1504} simulation. In the next section we will give predictions for the fuelling of star formation by recycled stellar ejecta in present-day central and satellite galaxies as a function of their halo and stellar mass. To be able to relate these predictions to observational diagnostics, we first explore the relation between recycling-fuelled star formation and present-day metallicity. We will show that the fact that metals are synthesized in stars and are distributed over the ISM as the evolving stellar populations lose mass, makes them an excellent observational proxy for the contribution of stellar ejecta to the SFR and stellar mass.

To study the \sfrratio, we only consider subhaloes with a non-zero\footnote{`Non-zero' means containing at least one star-forming gas particle, which corresponds to a specific SFR ($= \mathit{SFR} / M\sub{\ast}$) of $> 10^{-12}$ yr$^{-1}$ at $M\sub{\ast} \sim 10^{9}$ M$\sub{\odot}$ and $> 10^{-14}$ yr$^{-1}$ at $M\sub{\ast} \sim 10^{11}$ M$\sub{\odot}$.} SFR, while to study the \mstarratio, we only consider subhaloes with a non-zero stellar mass. For our fiducial simulation this yields samples of $44\,248$ and $325\,561$ subhaloes, respectively. In this section, however, we additionally require the subhaloes to have a galaxy stellar mass corresponding to at least $100$ gas particles, which yields samples of $14\,028$ and $16\,681$ subhaloes, respectively.

Fig.~\ref{fig:metallicity} shows the fraction of the SFR (left panel) and stellar mass (right panel) fuelled by recycling as a function of, respectively, the mass-weighted absolute metallicity $Z\sub{gas}$ of ISM gas (i.e. star-forming gas) and the mass-weighted absolute metallicity $Z\sub{\ast}$ of stars, both for present-day central galaxies\footnote{Although we do not explicitly show it, the results for central galaxies presented in this section are consistent with the results for satellite galaxies.}. We find strong correlations between these quantities, with more metal-rich galaxies having a larger fraction of their SFR and stellar mass contributed by recycling. The figure reveals tight power-law relations between \sfrratio and $Z\sub{gas}$, characterized by a Pearson correlation coefficient of $0.95$, and between \mstarratio and $Z\sub{\ast}$, with a correlation coefficient of $0.99$. For the former we find a $1\sigma$ scatter of $\sim 0.1 - 0.2$ dex for $Z\sub{gas} < 10^{-1.9}$ and $\lesssim 0.05$ dex for $Z\sub{gas} > 10^{-1.9}$, while for the latter we find an even smaller $1\sigma$ scatter of $\sim 0.01 - 0.03$ dex. Furthermore, as we show in Appendix~\ref{sec:resotests_metal}, both relations are converged with respect to the numerical resolution.

The tight relation between the contribution of recycled gas to star formation and metallicity is not surprising considering that heavy elements were produced in stars and that their abundance must therefore correlate with the importance of stellar ejecta as star formation fuel. The contribution of recycling to the stellar mass is equal to the ratio of the mean stellar metallicity ($\Big\langle Z\sub{\ast} \Big\rangle$) and the mean metallicity of the ejecta ($\Big\langle Z\sub{ej} \Big\rangle$) that were incorporated into the stars,
\begin{equation}\label{eq:mstar_metal_model}
\frac{M\sub{\ast,rec}}{M\sub{\ast}}=\frac{\Big\langle Z\sub{\ast} \Big\rangle}{\Big\langle Z\sub{ej} \Big\rangle}.
\end{equation}
The same holds for the contribution of stellar mass loss to the SFR,
\begin{equation}\label{eq:sfr_metal_model}
\frac{\mathit{SFR}\sub{rec}}{\mathit{SFR}}=\frac{\Big\langle Z\sub{gas} \Big\rangle}{\Big\langle Z\sub{ej} \Big\rangle}.
\end{equation}
The metallicity of the ejecta depends on the age and metallicity of the SSP, as well as on the IMF. From Fig.~\ref{fig:SSP_chabrier} we can see that for our (Chabrier) IMF, $\Big\langle Z\sub{ej} \Big\rangle \approx 0.033 / 0.45 \approx 0.073$ for a $10$ Gyr old SSP with solar metallicity. Hence, $\log\sub{10}$ \mstarratio $\approx \log\sub{10} Z\sub{\ast} + 1.1$, where the slope and normalization are close to the best-fit values that we determine below. Note that using ages of $100$ Myr and $10$ Myr instead of $10$ Gyr gives normalizations of $0.91$ and $0.77$, respectively. Using an age of $10$ Gyr but a stellar metallicity of $0.01$ Z$\sub{\odot}$ instead of Z$\sub{\odot}$ yields a normalization of $1.2$.

\begin{figure}
\begin{center}
\includegraphics[width=\columnwidth]{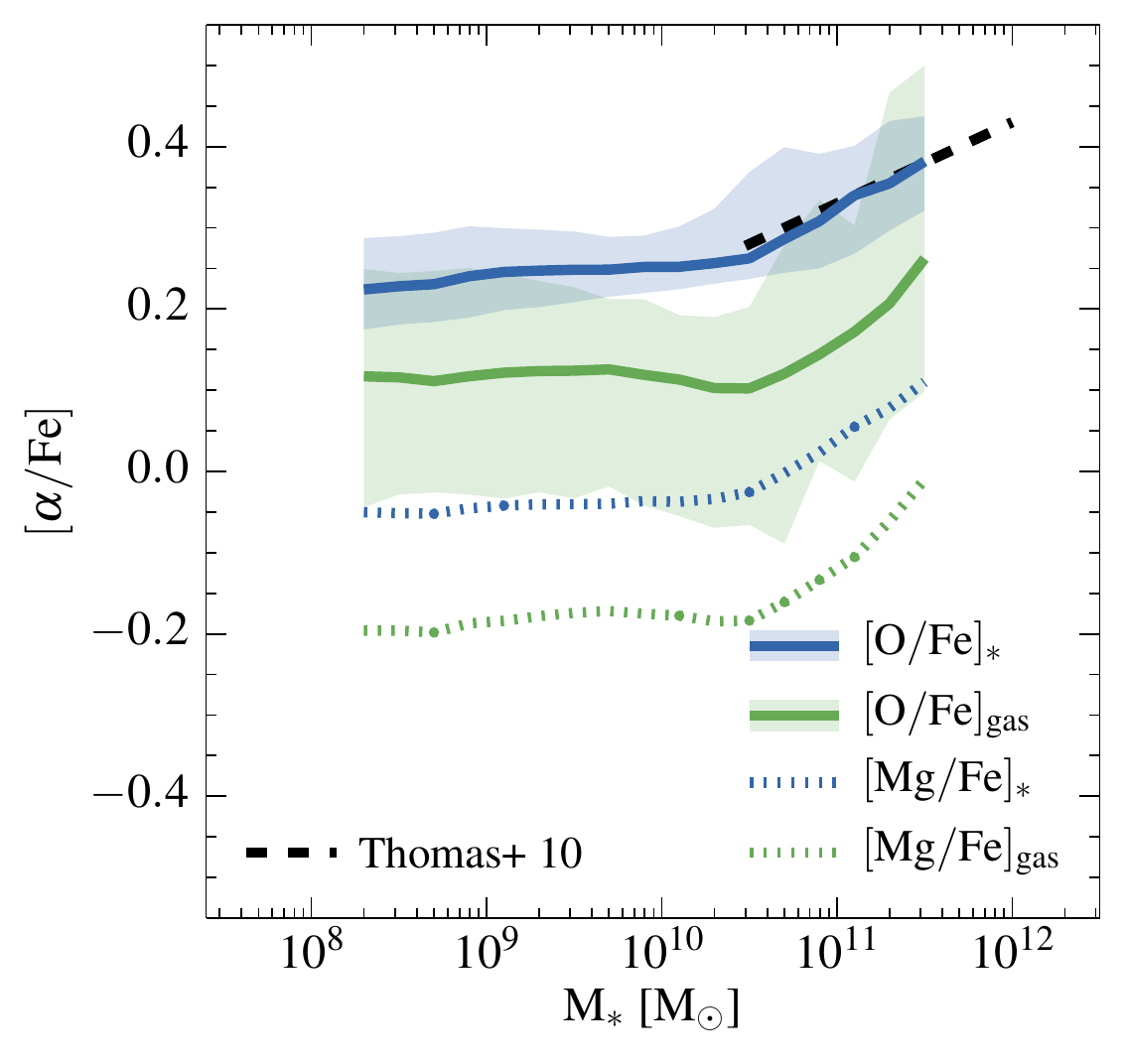}
\end{center}
\caption{The $\alpha$-element-to-iron abundance ratio of central galaxies at $z=0$ as a function of stellar mass. We show $[\alpha/\mathrm{Fe}]$, represented by $[\mathrm{O}/\mathrm{Fe}]$ (solid) and $[\mathrm{Mg}/\mathrm{Fe}]$ (dotted), of ISM gas (green) and stars (blue) as predicted by EAGLE, and compare with observations of the stellar $[\alpha/\mathrm{Fe}]$ from \citet{thomas_2010} (converted to a solar abundance ratio of $X\sub{\odot}\up{O}/X\sub{\odot}\up{Fe} = 4.44$). The curves show the median value in each logarithmic mass bin of size $0.2$ dex, if it contains at least $10$ haloes and the stellar mass corresponds to at least $100$ gas particles. The shaded regions mark the $10$th to $90$th percentiles, shown only for $[\mathrm{O}/\mathrm{Fe}]$. For $M\sub{\ast} \lesssim 10^{10.5}$ M$\sub{\odot}$, $[\mathrm{O}/\mathrm{Fe}]$ ($[\mathrm{Mg}/\mathrm{Fe}]$) is approximately constant at $\sim 0.1$ ($-0.2$) for gas and at $0.25$ ($-0.05$) for stars. For $M\sub{\ast} \gtrsim 10^{10.5}$ M$\sub{\odot}$, $[\mathrm{O}/\mathrm{Fe}]$ and $[\mathrm{Mg}/\mathrm{Fe}]$ increase with stellar mass, in such a way that the slope matches the observations, reflecting the enhancement in the contribution to the SFR and stellar mass from massive stars relative to that from intermediate-mass stars.}
\label{fig:ab_ratio}
\end{figure}

There is, however, an additional factor at play that may distort the one-to-one correlation between the contribution of recycled gas to the SFR (and therefore to the stellar mass) and metallicity, namely the relative significance of the different mass loss channels. This depends on the timescale on which stars are formed, but is also affected by processes like stellar and AGN feedback. Given that the ejecta from massive stars have $\sim 4 - 6$ times higher metallicity than those from intermediate-mass stars (dependent on metallicity; see Fig.~\ref{fig:SSP_chabrier}), a higher contribution of the mass loss from massive stars to the SFR (for fixed \sfrratio) would yield a higher ISM metallicity, and would hence change the relation between \sfrratio and $Z\sub{gas}$. As we will show in Section~\ref{sec:massdep_AGB_SN}, the contribution to the SFR of the mass loss from massive stars relative to that from AGB stars varies as a function of stellar mass, and in particular increases at the high-mass end. This introduces a mild mass dependence in the \sfrratio - $Z\sub{gas}$ and \mstarratio - $Z\sub{\ast}$ relations\footnote{Another factor is that the metal yields depend on metallicity (Fig.~\ref{fig:SSP_chabrier}). This can change the \sfrratio - $Z\sub{gas}$ and \mstarratio - $Z\sub{\ast}$ relations even if the contributions from the different channels remain fixed. However, even a factor of $100$ variation in the metallicity changes the metallicity of the stellar ejecta by only a few percent, which is significantly smaller than the effect of the change in the relative channel contributions in massive galaxies.}. In order to relate this variation of the relative contribution from different mass loss channels to an observational diagnostic, we consider the average $\alpha$-enhancement, $[\alpha/\mathrm{Fe}]$, represented by $[\mathrm{O}/\mathrm{Fe}]$ (as oxygen dominates the $\alpha$-elements in terms of mass fraction), of ISM gas and stars. The fact that $\alpha$-elements are predominantly synthesized in massive stars, whereas of iron $\sim 50\%$ is contributed by intermediate-mass stars in the form of SN Type Ia explosions and winds from AGB stars \citep[e.g.][]{wiersma_2009b}, makes $[\alpha/\mathrm{Fe}]$ a good tracer for the relative importance of massive stars.

Adopting the usual definition of the abundance ratio,
\begin{equation}\label{eq:ab_ratio}
\left[\frac{\mathrm{O}}{\mathrm{Fe}}\right] = \log\sub{10} \left(\frac{X\up{O}}{X\up{Fe}}\right) - \log\sub{10} \left(\frac{X\sub{\odot}\up{O}}{X\sub{\odot}\up{Fe}}\right),
\end{equation}
where $X^x$ is the mass fraction of element $x$ and $X\sub{\odot}\up{O}/X\sub{\odot}\up{Fe} = 4.44$ is the solar abundance ratio \citep{asplund_2009}, we show $[\mathrm{O}/\mathrm{Fe}]$ as a function of stellar mass in Fig.~\ref{fig:ab_ratio}. The curves show the median in logarithmic mass bins of size $0.2$ dex that contain at least $10$ haloes and correspond to a stellar mass of at least $100$ gas particles. The shaded regions mark the $10$th to $90$th percentile ranges. In both the gas-phase (green, solid) and the stellar phase (blue, solid), $[\mathrm{O}/\mathrm{Fe}]$ is approximately constant at $\sim 0.1$ and $\sim 0.25$, respectively, for $M\sub{\ast} \lesssim 10^{10.5}$ M$\sub{\odot}$, but increases with stellar mass for $M\sub{\ast} \gtrsim 10^{10.5}$ M$\sub{\odot}$. Comparing this to observations of the stellar $[\alpha/\mathrm{Fe}]$ for a sample of $3360$ early-type galaxies from \citet{thomas_2010} (best-fit relation, after correcting for the difference in the set of solar abundances used; black dashed line), we find excellent agreement in terms of the slope and the normalization. While this is encouraging, suggesting that we capture the right mass dependence in the \sfrratio - $Z\sub{gas}$ and \mstarratio - $Z\sub{\ast}$ relations and that the cooling rates \citep[which are dominated by oxygen at $T \sim 2 \times 10^5$ K and by iron at $T \sim 10^6$ K; see][]{wiersma_2009a} employed by the simulation are realistic, the predicted abundance ratio is uncertain by a factor of $> 2$ due to uncertainties in the nucleosynthetic yields and SN Type Ia rate \citep{wiersma_2009b}. It is therefore somewhat surprising that the agreement in the normalization is this good. If we consider $[\mathrm{Mg}/\mathrm{Fe}]$, which is another indicator of $[\alpha/\mathrm{Fe}]$ often used in the literature, of ISM gas (blue, dotted) and stars (green, dotted), we find an offset of $\sim 0.3$ dex with respect to the observed $[\alpha/\mathrm{Fe}]$. Note that the size of this offset is dependent on the adopted set of solar abundances. The slope, on the other hand, still matches the observed one, implying that the offset can be attributed to a constant uncertainty factor in the (massive star) yields.

Motivated by the tight power-law relations shown in Fig.~\ref{fig:metallicity}, we fit the relation between the recycled gas contribution to the SFR and ISM metallicity with the following function, including a term describing the variation in the relative channel contributions:
\begin{equation}\label{eq:sfr_metal}
\log\sub{10} \frac{\mathit{SFR}\sub{rec}}{\mathit{SFR}} = 0.87 \log\sub{10} Z\sub{gas} - 0.40 \left[\frac{\mathrm{O}}{\mathrm{Fe}}\right]\sub{gas} + 0.90,
\end{equation}
where the values of the three free parameters have been obtained using least square fitting. Note that the metallicity $Z$ is the average mass fraction of metals and is thus independent of the adopted solar value. Similarly, we determine the best-fit relation between the recycled gas contribution to the stellar mass and stellar metallicity:
\begin{equation}\label{eq:mstar_metal}
\log\sub{10} \frac{M\sub{\ast,rec}}{M\sub{\ast}} = 0.91 \log\sub{10} Z\sub{\ast} - 0.28 \left[\frac{\mathrm{O}}{\mathrm{Fe}}\right]\sub{\ast} + 0.92.
\end{equation}
We show these relations in Fig.~\ref{fig:metallicity} for galaxies with $M\sub{\ast} \sim 10^{9.5}$ M$\sub{\odot}$ (red, solid line), $M\sub{\ast} \sim 10^{10.5}$ M$\sub{\odot}$ (blue, dot-dashed line) and $M\sub{\ast} \sim 10^{11.5}$ M$\sub{\odot}$ (purple, dashed line), where we use the median values of $[\alpha/\mathrm{Fe}]\sub{gas}$ and $[\alpha/\mathrm{Fe}]\sub{\ast}$ in stellar mass bins of $0.2$ dex centred on the respective masses. As expected from Fig.~\ref{fig:ab_ratio}, the relations at $M\sub{\ast} \sim 10^{9.5}$ M$\sub{\odot}$ and $M\sub{\ast} \sim 10^{10.5}$ M$\sub{\odot}$ are consistent, as a result of the median $[\alpha/\mathrm{Fe}]$ (of gas and stars) being constant for $M\sub{\ast} \lesssim 10^{10.5}$ M$\sub{\odot}$. On the other hand, galaxies with $M\sub{\ast} \sim 10^{11.5}$ M$\sub{\odot}$ have \sfrratio and \mstarratio that are $\sim 0.06$ and $\sim 0.04$ dex \emph{lower} at fixed metallicity due to an enhancement in the contribution from massive stars relative to that from intermediate-mass stars (reflected by their enhanced $[\alpha/\mathrm{Fe}]$ abundance ratio). These offsets are somewhat larger than the $1\sigma$ scatter in the relation for all galaxy masses (which is set by the scatter at $M\sub{\ast} < 10^{10}$ M$\sub{\odot}$), indicating that the variation of the channel contributions at $M\sub{\ast} > 10^{10.5}$ M$\sub{\odot}$ significantly impacts upon the relation between metallicity and recycling-fuelled star formation in high-mass galaxies. It leads to a reduction of \sfrratio and \mstarratio at fixed metallicity that increases with stellar mass, and will therefore make any turnover or flattening at the high-mass end of the relation between recycled gas contributions and stellar mass \citep[as seen in the mass-metallicity relation; see][]{tremonti_2004,gallazzi_2005,kewley+ellison_2008,andrews+martini_2013,zahid_2014b} more pronounced. We demonstrate the useful link that equations~(\ref{eq:sfr_metal}) and (\ref{eq:mstar_metal}) provide between the importance of gas recycling and observational diagnostics in Section~\ref{sec:massdep_eagle}.

We note that the parameters of equations~(\ref{eq:sfr_metal}) and (\ref{eq:mstar_metal}) are insensitive to the specific implementation of subgrid processes like star formation, stellar feedback and AGN feedback\footnote{The adopted IMF is an exception, as it determines the mass and metallicity of gas returned by stellar populations, as well as the relative contribution from massive stars with respect to intermediate-mass stars.}, as for EAGLE changing their implementation affects the recycled gas contributions and metallicities in a similar way. Note that this may not be true if the metallicity of galactic winds differs significantly from the metallicity of the ISM, as might for example happen if metals are preferentially ejected \citep[e.g.][]{maclow+ferrara_1999,creasey_2015}, or if instead galactic winds are metal-depressed \citep[e.g.][]{zahid_2014a}.


\subsection{Dependence on halo and galaxy mass at $z=0$}
\label{sec:massdep_eagle}

In this section we investigate how the fractional contribution of recycled gas to the present-day SFR and stellar mass of galaxies depends on their halo and stellar mass. Note that, because of the tight relation with metallicity that we established in Section~\ref{sec:metallicity}, many conclusions that we draw here carry over to the mass-metallicity relation. We study both central (Section~\ref{sec:massdep_cen}) and satellite (Section~\ref{sec:massdep_sat}) galaxies and, in addition to the total contribution of gas recycling, assess the relative significance of the different mass loss channels (Section~\ref{sec:massdep_AGB_SN}). We also briefly explore how fuelling by gas recycling depends on the distance from the galactic centre (Section~\ref{sec:massdep_radius}). While we mainly present results from our fiducial \emph{Ref-L100N1504} simulation, we also show a brief comparison with the results from \emph{Recal-L025N0752} for central galaxies.

\subsubsection{Gas recycling in central galaxies}
\label{sec:massdep_cen}

\begin{figure*}
\begin{center}
\includegraphics[width=0.8\textwidth]{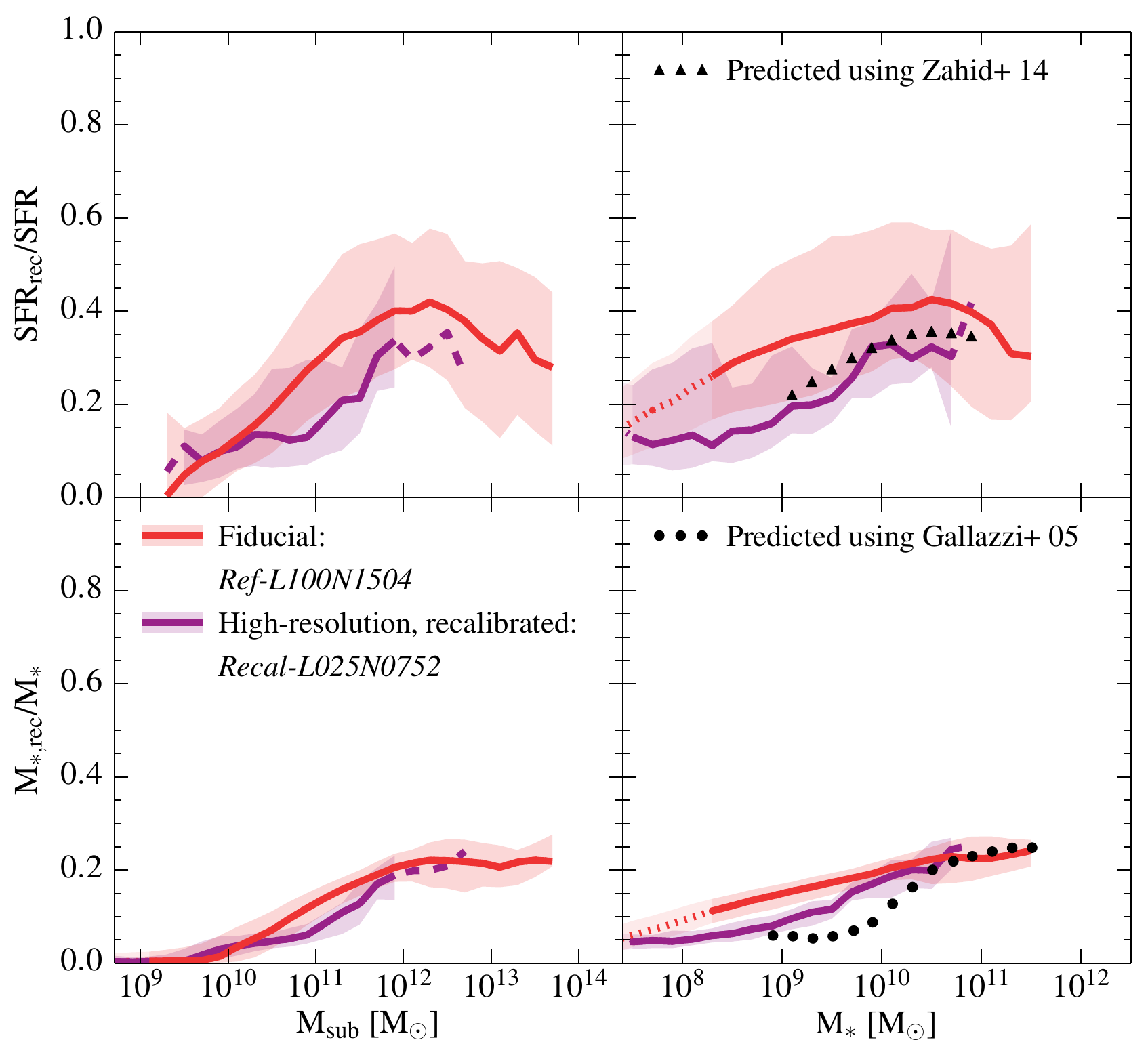}
\end{center}
\caption{The fractional contribution of gas recycled from stellar mass loss to the SFR (top) and stellar mass (bottom) of central galaxies at $z=0$ as a function of their subhalo mass (left) and stellar mass (right). We show the results for the fiducial EAGLE model (\emph{Ref-L100N1504}; red) and the high-resolution, recalibrated model (\emph{Recal-L025N0752}; purple). We only consider subhaloes with a non-zero SFR (top panels) or a non-zero stellar mass (bottom panels). The curves show the median value in each logarithmic mass bin of size 0.2 dex, if it contains at least $10$ galaxies. The shaded regions mark the $10$th to $90$th percentile ranges. The solid curves become dotted when the subhalo (stellar) mass corresponds to fewer than $100$ dark matter (baryonic) particles and become dashed (for \emph{Recal-L025N0752} only) when there are less than $10$ haloes per bin.The contribution of recycled gas to the SFR and stellar mass first increases with mass, turns over at $M\sub{\ast} \sim 10^{10.5}$ M$\sub{\odot}$ ($M\sub{sub} \sim 10^{12.2}$ M$\sub{\odot}$), and then decreases or remains constant at higher mass. This trend is regulated by the efficiency of the feedback from star formation (AGN) at low (high) masses: galactic winds eject gas from the ISM, where stellar mass loss accumulates, and therefore preferentially reduce the SFR and stellar mass contributed by recycling. The black points represent our best estimate of the recycled gas contributions to the SFR and stellar mass (for a central galaxy with a Milky Way-like mass: $35 \%$ and $20 \%$, respectively), calculated by applying equations~(\ref{eq:sfr_metal}) and (\ref{eq:mstar_metal}) to the observed gas-phase metallicities from \citet{zahid_2014b} and the observed stellar metallicities from \citet{gallazzi_2005}. For $M\sub{\ast} \gtrsim 10^{10}$ M$\sub{\odot}$, these estimates agree to better than a factor of $\sim 1.6$ ($0.2$ dex) with the median predictions computed directly from EAGLE.}
\label{fig:massdep_eagle_reso}
\end{figure*}

Fig.~\ref{fig:massdep_eagle_reso} shows the contribution of recycled stellar mass loss to the present-day SFR and stellar mass of central galaxies as a function of their mass in the \emph{Ref-L100N1504} (red) and \emph{Recal-L025N0752} (purple) simulations. We plot \sfrratio in the top row and \mstarratio in the bottom row as a function of subhalo mass (left column) and stellar mass (right column). Focusing first on the fiducial \emph{Ref-L100N1504} simulation, the general trend in all four panels is that, at masses $M\sub{sub} \lesssim 10^{12.2}$ M$\sub{\odot}$ or $M\sub{\ast} \lesssim 10^{10.5}$ M$\sub{\odot}$, the fraction of the SFR and stellar mass contributed by recycling increases with mass. This is the regime where the greater depth of the gravitational potential well, as well as the higher pressure and density of the ISM and CGM, towards higher masses, make it harder for feedback (dominated by star formation) to eject gas from the galaxy. As we will show explicitly with a model comparison in Section~\ref{sec:massdep_owls}, a reduced efficiency of stellar feedback at driving galactic outflows enhances the contribution from recycled gas to the SFR and stellar mass. This can be understood by considering that these winds (if stellar feedback is efficient) are launched from the dense star-forming regions with relatively high abundances of gas from stellar mass loss. Hence, more efficient winds will preferentially reduce $\mathit{SFR}\sub{rec}$ with respect to the total $\mathit{SFR}$ (thereby reducing \sfrratio), whereas in the case of less efficient winds this effect will be less (thereby enhancing \sfrratio).

At the high-mass end, \sfrratio and \mstarratio turn over at $M\sub{\ast} \sim 10^{10.5}$ M$\sub{\odot}$ ($M\sub{sub} \sim 10^{12.2}$ M$\sub{\odot}$), and then decrease and remain constant, respectively, at higher masses. In this mass regime, the trend is regulated by the efficiency of the feedback from AGN, which becomes stronger in more massive systems. Even though this type of feedback is not associated with any replenishment of the ISM gas reservoir (as opposed to feedback from star formation, which directly provides the gas for recycling), it does have a significant impact on the rate at which galaxies consume the enriched ISM gas. If AGN are efficient at launching galactic outflows, they preferentially remove or disperse the dense ISM gas from the central regions, in which the abundance of stellar ejecta is high, thereby reducing \sfrratio and \mstarratio.

We note that while AGN feedback regulates the turnover at $M\sub{\ast} \sim 10^{10.5}$ M$\sub{\odot}$, the use of the $30$ pkpc 3D aperture also plays a role in shaping the behaviour of \sfrratio and \mstarratio at the high-mass end. As we show in Appendix~\ref{sec:aperture}, \sfrratio and \mstarratio at $M\sub{\ast} \gtrsim 10^{10.5}$ M$\sub{\odot}$ ($M\sub{sub} \gtrsim 10^{12.2}$ M$\sub{\odot}$) are somewhat enhanced, and their slopes become somewhat shallower, if an aperture is applied. This is consistent with the fraction of the SFR  and stellar mass fuelled by recycled gas being larger in the inner parts of galaxies (see Fig.~\ref{fig:massdep_radius}). Without an aperture, \mstarratio decreases with halo and galaxy mass (similar to \sfrratio), instead of remaining roughly constant if an aperture is applied.

At the mass scale of the turnover, the fractional contribution of recycled gas to the SFR is at a maximum. A galaxy of this mass, $M\sub{\ast} \sim 10^{10.5}$ M$\sub{\odot}$, is too massive to have effective star formation-driven outflows but still too small for AGN feedback to be effective. Not surprisingly, this mass scale coincides with the peak in the galaxy formation efficiency (see fig. 8 of S15), which is consistent with the efficiency of feedback being the main driver of \sfrratio and \mstarratio. The fiducial EAGLE model indicates that for a Milky Way-like galaxy, which is at the peak of the galaxy formation efficiency, $40 \%$ of its present-day SFR and $20 \%$ of its present-day stellar mass is due to the recycling of stellar mass loss.

Because of the tight correlation between recycling-fuelled star formation and metallicity, our findings have direct implications for the origin of the mass-metallicity relations for   ISM gas and stars. They imply that the increase in metallicity with stellar mass at $M\sub{\ast} \lesssim 10^{10.5}$ M$\sub{\odot}$ is due to the decreasing efficiency of stellar feedback at driving galactic outflows, while the shape at higher mass is governed by the efficiency of AGN feedback \citep[see also][for discussion on the relation between feedback and metallicity]{zahid_2014a,peeples_2014,creasey_2015}. Conversely, the difference between the \emph{Ref-L100N1504} and \emph{Recal-L025N0752} simulations in Fig.~\ref{fig:massdep_eagle_reso}, as well as their (expected) agreement with observations, should mimic the results for the mass-metallicity relation (see fig. 13 of S15). Indeed, while \emph{Ref-L100N1504} and \emph{Recal-L025N0752} yield similar trends, they do differ quantitatively by a factor of $\sim 2$ ($0.3$ dex) in \sfrratio and \mstarratio at $M\sub{\ast} \sim 10^{9}$ M$\sub{\odot}$ ($M\sub{sub} \sim 10^{11}$ M$\sub{\odot}$). This difference decreases towards higher masses, where for $M\sub{\ast} \gtrsim 10^{9.8}$ M$\sub{\odot}$ ($M\sub{sub} \gtrsim 10^{11.6}$ M$\sub{\odot}$), \emph{Ref-L100N1504} and \emph{Recal-L025N0752} are converged in terms of \mstarratio and broadly consistent in terms of \sfrratio (considering the substantial amount of scatter and relatively poor sampling of the high-mass regime by the \emph{Recal-L025N0752} model). S15 showed that for $M\sub{\ast} \gtrsim 10^{9.8}$ M$\sub{\odot}$, the metallicities of galaxies in \emph{Ref-L100N1504} and \emph{Recal-L025N0752} agree with the observations equally well. They agree with the observed gas-phase metallicities from \citet{zahid_2014b} to better than $0.1$ dex and with \citet{tremonti_2004} to better than $0.2$ dex, and with the observed stellar metallicities from \citet{gallazzi_2005} to within the observational uncertainties (which are $> 0.5$ dex at $M\sub{\ast} < 10^{10}$ M$\sub{\odot}$ and smaller at higher masses). For $M\sub{\ast} \lesssim 10^{9.8}$ M$\sub{\odot}$, on the other hand, the metallicities of galaxies in \emph{Recal-L025N0752} are in better agreement with the observations, from which we conclude that the values of \sfrratio and \mstarratio predicted by \emph{Recal-L025N0752} are more reliable than those predicted by the fiducial model. Note, however, that the large systematic uncertainties associated with the calibration of the diagnostics prevent any strong conclusions. In order to limit the number of model curves plotted in each figure, from here on we only plot the results from \emph{Ref-L100N1504} and ask the reader to keep in mind the slight overprediction of \sfrratio and \mstarratio at $M\sub{\ast} \lesssim 10^{9.8}$ M$\sub{\odot}$.

Finally, in contrast to the predictions computed directly from EAGLE, which at low masses depend on the adopted numerical resolution, the relations between gas recycling and metallicity given in equations~(\ref{eq:sfr_metal}) and (\ref{eq:mstar_metal}) provide a way of estimating \sfrratio and \mstarratio, that is independent of the resolution. Moreover, these relations are insensitive to the subgrid models for feedback. We apply the relations to the observed mass-metallicity relations from \citet{zahid_2014b} and \citet{gallazzi_2005}, using the median $[\mathrm{O}/\mathrm{Fe}]$ from EAGLE in each stellar mass bin, to estimate \sfrratio (triangular points, upper-right panel of Fig.~\ref{fig:massdep_eagle_reso}) and \mstarratio (circular points, lower-right panel of Fig.~\ref{fig:massdep_eagle_reso}) as a function of stellar mass. These estimates agree qualitatively with \sfrratio and \mstarratio computed directly from the fiducial EAGLE model, showing a steep increase with mass for $M\sub{\ast} \lesssim 10^{10.5}$ M$\sub{\odot}$, followed by turnover and even a slight downturn in \sfrratio at higher masses\footnote{Note that, even though the mass-metallicity relation observed by \citet{zahid_2014b} does not exhibit a decrease in the metallicity at the high-mass end, the recycled gas contribution to the SFR can still show a slight downturn, due to the change in the relative contributions from the different mass loss channels (as discussed in Section~\ref{sec:metallicity}).}. Quantitatively, the black points are in good agreement with the fiducial EAGLE model for $M\sub{\ast} \gtrsim 10^{10}$ M$\sub{\odot}$ and with \emph{Recal-L025N0752} also at lower masses, as expected from the comparison of the mass-metallicity relation with the observations presented in S15. If the discrepancy between the predicted and observed mass-metallicity relation exceeds the systematic error due to calibration uncertainties in the observations, then the black points represent our best estimates of the recycled gas contributions to the SFR and stellar mass. For a Milky Way-like galaxy ($M\sub{\ast} \sim 10^{10.5}$ M$\sub{\odot}$), we find these contributions to be $35 \%$ and $20 \%$, respectively.

\subsubsection{Gas recycling in satellite galaxies}
\label{sec:massdep_sat}

\begin{figure*}
\begin{center}
\includegraphics[width=0.8\textwidth]{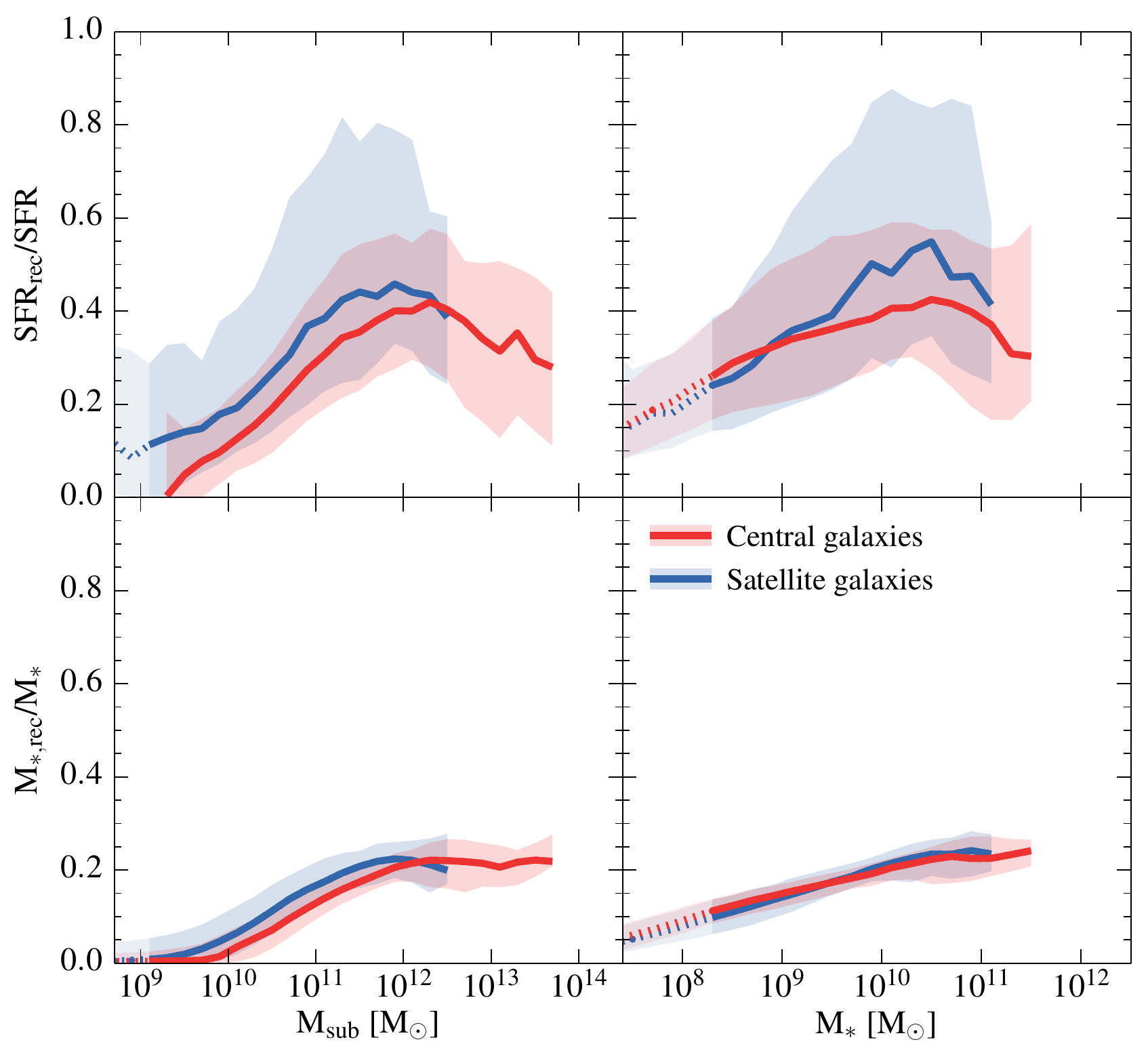}
\end{center}
\caption{As Fig.~\ref{fig:massdep_eagle_reso}, but showing the results for central (red) and satellite (blue) galaxies from the fiducial EAGLE model. The recycled gas contributions to the SFR and stellar mass in satellites are broadly consistent with the ones in similarly massive centrals, since the efficiency of stellar and AGN feedback is the controlling factor in fuelling star formation with recycled gas. However, in the inefficient feedback regime ($M\sub{\ast} \sim 10^{10.5}$ M$\sub{\odot}$), satellites with low gas fractions can reach recycling-fuelled SFR fractions as high as $\sim 90 \%$, with a median that exceeds the one in similarly massive centrals (see also Fig.~\ref{fig:sat_massbin}).}
\label{fig:massdep_eagle}
\end{figure*}

Having studied the recycling-fuelled star formation in present-day central galaxies, we now compare these with the results for present-day satellite galaxies. Fig.~\ref{fig:massdep_eagle} shows the SFR and stellar mass contributed by recycling for both central (red; as in Fig.~\ref{fig:massdep_eagle_reso}) and satellite (blue) galaxies, as predicted by the fiducial \emph{Ref-L100N1504} simulation. In general, these are broadly similar for centrals and satellites. However, we identify two important differences. Firstly, in the left panels, where we show the two ratios as a function of subhalo mass, the relations for satellite galaxies are shifted towards lower masses relative to those for central galaxies. Satellites lose a fraction of their dark matter subhalo mass (but less stellar mass) upon infall onto the group dark matter halo as a result of tidal stripping. Hence, this shift illustrates the fact that satellite galaxies live in smaller (sub)haloes than central galaxies of similar stellar mass. Secondly, in the top-right panel, at a mass scale of $M\sub{\ast} \sim 10^{10.5}$ M$\sub{\odot}$, satellites show significantly higher \sfrratio (with a median of $\sim 0.5$ and a $90$th percentile value of $\sim 0.85$) than centrals, whereas at lower and higher masses this difference is smaller. Hence, in the regime where both stellar feedback and AGN feedback are relatively inefficient, gas recycling plays a more important role in fuelling ongoing star formation in satellite galaxies than in central galaxies. For \mstarratio, on the other hand, there is no difference between centrals and satellites, because satellites formed the majority of their stars while they were still centrals.

\begin{figure*}
\begin{center}
\includegraphics[width=0.49\textwidth]{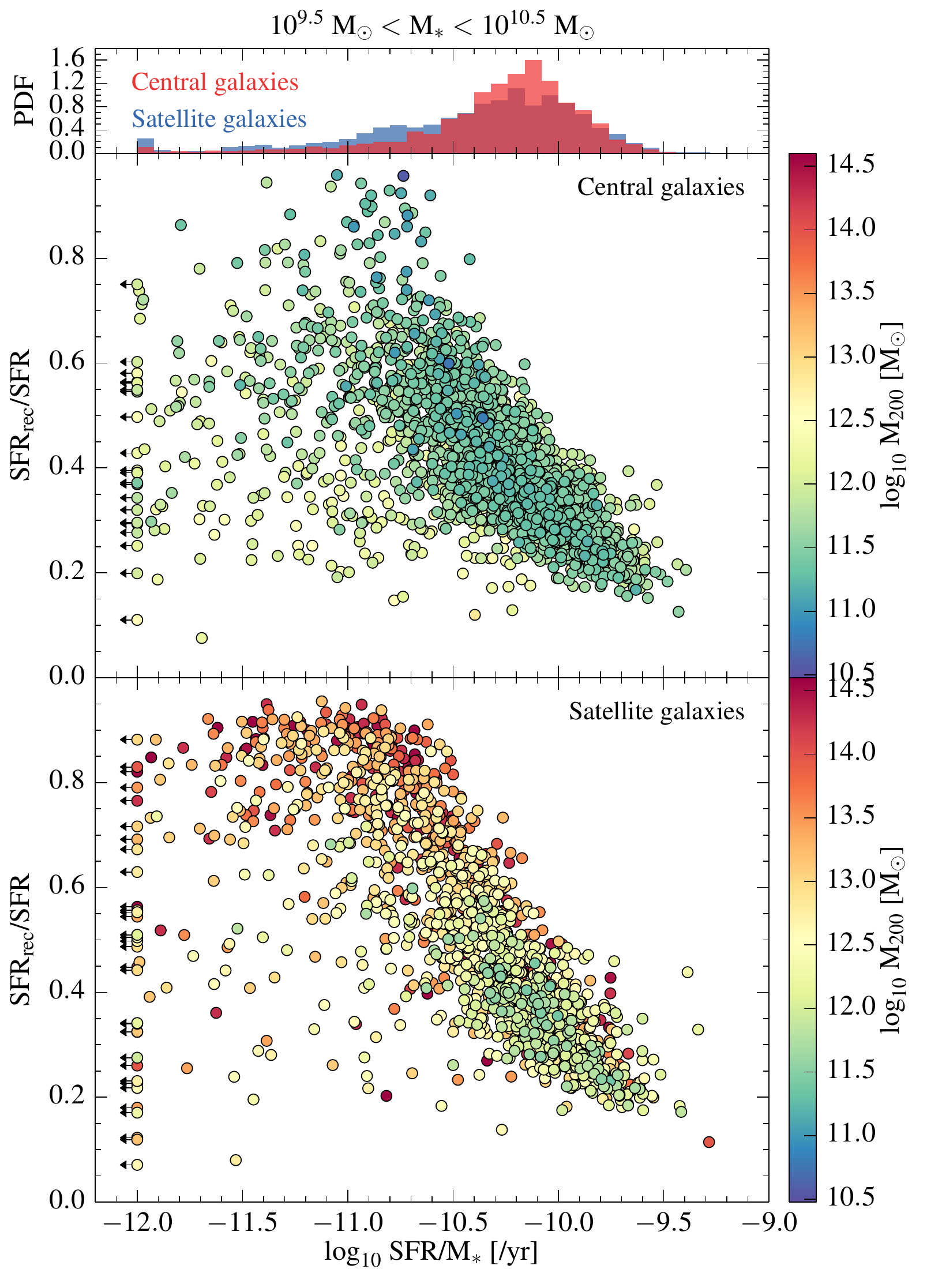}
\includegraphics[width=0.49\textwidth]{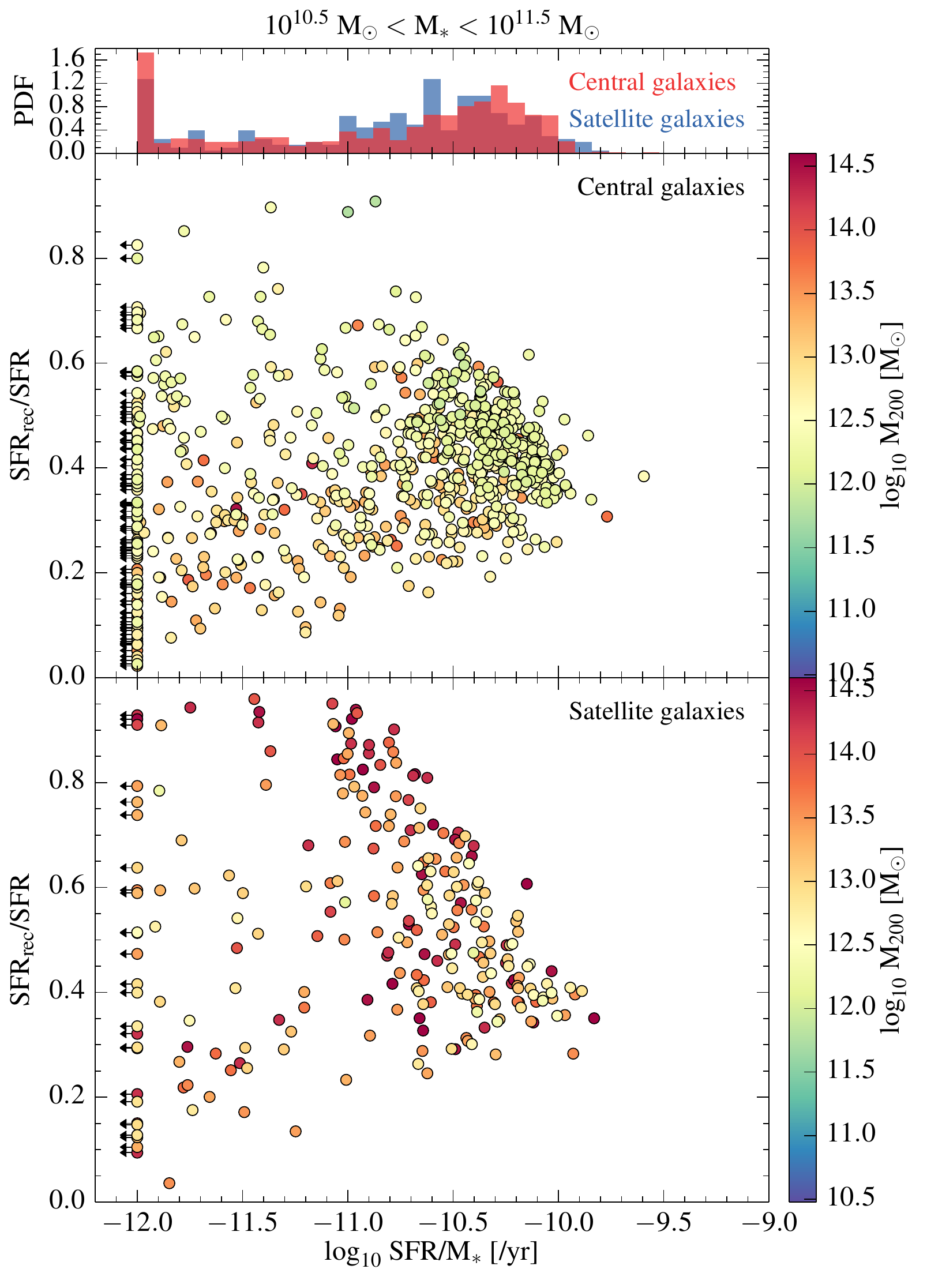}
\end{center}
\caption{The SFR fuelled by recycling as a function of sSFR ($= \mathit{SFR}/M\sub{\ast}$), colour-coded by host halo mass ($M\sub{200}$), for central galaxies (upper panels) and satellite galaxies (lower panels) with stellar masses $10^{9.5}$ M$\sub{\odot} < M\sub{\ast} < 10^{10.5}$ M$\sub{\odot}$ (left) and $10^{10.5}$ M$\sub{\odot} < M\sub{\ast} < 10^{11.5}$ M$\sub{\odot}$ (right). The histograms at the top show the distributions of the sSFR for centrals and satellites in these two mass bins. Galaxies with $\mathit{SFR} / M\sub{\ast} < 10^{-12}$ yr$^{-1}$ are shown as upper limits. For the centrals, the relation between the recycling-fuelled SFR and the sSFR changes from an anticorrelation at lower mass, which is a result of the tight relation with ISM gas fraction, to a (weak) correlation at higher mass, which is driven by AGN feedback. The satellites, on the other hand, show a similar behaviour, but retain in both mass ranges a large population of low-sSFR galaxies that rely heavily on stellar mass loss for fuelling ongoing star formation (contributing $\gtrsim 90 \%$).}
\label{fig:sat_massbin}
\end{figure*}

To get a better understanding of the difference between centrals and satellites, we consider the relation between \sfrratio and specific SFR ($= \mathit{SFR}/M\sub{\ast}$, sSFR). Fig.~\ref{fig:sat_massbin} shows this relation for centrals (upper panels) and satellites (lower panels) with masses $10^{9.5}$ M$\sub{\odot} < M\sub{\ast} < 10^{10.5}$ M$\sub{\odot}$ (left) and $10^{10.5}$ M$\sub{\odot} < M\sub{\ast} < 10^{11.5}$ M$\sub{\odot}$ (right), where the histograms at the top compare the distributions of sSFRs. In order to limit the dynamical range plotted, galaxies with $\mathit{SFR} / M\sub{\ast} < 10^{-12}$ yr$^{-1}$ are shown as upper limits. The colour coding indicates the mass of the parent dark matter halo, $M\sub{200}$, in which these galaxies reside. For centrals, $M\sub{200}$ is generally closely related to the mass of the subhalo and the stellar mass, whereas for satellites the mass of the host halo they have fallen onto is only weakly related to their own mass. For satellites, $M\sub{200}$ instead serves as a proxy for the strength of any environmental effects like ram-pressure stripping \citep{gunn_gott_1972} or strangulation \citep{larson_1980}.

Focusing first on the centrals in the lower stellar mass bin, we see a clear anticorrelation between the fraction of the SFR that is fuelled by recycling and the sSFR. This can be explained by the fact that sSFR is closely related to the gas fraction ($= M\sub{\mathrm{gas}} / (M\sub{\mathrm{gas}} + M\sub{\ast})$ with $M\sub{gas}$ the ISM mass): higher gas fractions generally correspond to higher sSFRs. Also, since the ISM is comprised of both processed and unprocessed gas, whereas stellar mass only provides enriched gas for recycling, an enhanced gas fraction typically implies that a greater fraction of the star formation in the ISM is fuelled by unprocessed, `nonrecycled' gas (i.e. \sfrratio is low). On the other hand, galaxies with a low sSFR and a correspondingly low gas fraction will have a higher fraction of their SFR contributed by recycling. Considering the centrals in the higher mass bin, we see that while part of this relation is still in place, a significant fraction lies away from the main relation towards the lower left. This is the result of efficient AGN feedback, which suppresses both the sSFR (`quenching') and the \sfrratio of galaxies, as AGN feedback is more important at higher halo masses. This explanation is consistent with the enhanced halo masses of the galaxies in this regime ($M\sub{200} \sim 10^{13}$ M$\sub{\odot}$ for centrals with $\mathit{SFR} / M\sub{\ast} \sim 10^{-11.5}$ yr$^{-1}$ and \sfrratio $\sim 0.2$, compared to $M\sub{200} \sim 10^{12}$ M$\sub{\odot}$ for centrals with $\mathit{SFR} / M\sub{\ast} \sim 10^{-10.5}$ yr$^{-1}$ and \sfrratio $\sim 0.5$). At even higher central galaxy mass scales than shown in Fig.~\ref{fig:sat_massbin}, the anticorrelation between sSFR and \sfrratio disappears entirely. Instead, the relation transforms into an AGN feedback-controlled \emph{correlation} (although weak due to small number statistics, with a Pearson correlation coefficient of $0.43$ for $M\sub{\ast} > 10^{11.0}$ M$\sub{\odot}$), such that the galaxies with the lowest sSFRs have the lowest recycling-fuelled SFRs.

Having investigated the mechanism driving the sSFR - \sfrratio trends in the two mass regimes for centrals, we now consider satellites. In the lower mass bin the anticorrelation between sSFR and \sfrratio is similar to that for centrals, although the histogram at the top shows that the sSFR distribution for satellites has larger scatter towards low sSFRs. This corresponds to a population of satellite galaxies at $\mathit{SFR} / M\sub{\ast} \sim 10^{-11}$ yr$^{-1}$ with fractional contributions of recycled gas to their SFR as high as $90 - 95 \%$. This high-\sfrratio regime is more frequently populated by satellites than similarly massive centrals, reflecting the substantially greater scatter in the satellite curves towards high values of \sfrratio at $M\sub{\ast} \sim 10^{10.5}$ seen in Fig.~\ref{fig:massdep_eagle}. As indicated by the colour coding in Fig.~\ref{fig:sat_massbin}, this population of satellites is hosted by relatively massive group dark matter haloes ($M\sub{200} \sim 10^{14}$ M$\sub{\odot}$), implying that their low sSFRs and gas fractions are the result of the cessation of fresh gas infall (either because cooling is inefficient or because the satellite's hot gas reservoir was stripped), and/or a (partial) removal of cold gas from the disc \citep[see e.g.][]{bahe+mccarthy_2015,mistani_2015}. Both scenarios lead to a depletion of the ISM gas reservoir and a greater dependence on stellar mass loss for replenishing it.

Finally, we focus on the satellite galaxies in the higher mass bin, shown in the bottom-right of Fig.~\ref{fig:sat_massbin}. Whereas most similarly massive central galaxies that have moved away from the sSFR - \sfrratio anticorrelation, have moved towards low sSFR and low \sfrratio under the influence of AGN feedback, there is still a significant population of satellite galaxies occupying the high-\sfrratio region. Inspecting the masses of the BHs residing in these satellites (not shown) we see that they are significantly lower than the masses of BHs in centrals of similar stellar mass. We infer that this is again due to the depletion of the satellite ISM gas, thereby preventing efficient BH growth. This explains why the AGN feedback in these satellites is unable to suppress the recycled gas contribution to the SFR.

We conclude that the SFR and stellar mass contributed by recycling are broadly consistent between central and satellite galaxies over a wide range of galaxy masses, because gas recycling is governed primarily by the efficiency of stellar and AGN feedback. However, in satellites with a stellar mass similar to that of the Milky Way, the mass scale at which feedback is least efficient at suppressing star formation, the recycled gas contribution to the SFR often exceeds the one in similarly massive centrals (and can even reach $\gtrsim 90\%$), as the depletion of their ISM gas reservoir makes them more reliant on stellar mass loss for fuelling ongoing star formation.

Our findings are consistent with the observational inference that, at a given stellar mass, satellites are more metal-rich than centrals \citep{pasquali_2012,peng+maiolino_2014}. We explain the origin of their different mass-metallicity relation as a consequence of satellites being subject to environmental processes like ram-pressure stripping and strangulation, which prevent the dilution of the ISM reservoir by metal-poor gas.

\subsubsection{Contributions from AGB stars, SN Type Ia and massive stars}
\label{sec:massdep_AGB_SN}

\begin{figure*}
\begin{center}
\includegraphics[width=0.9123\textwidth]{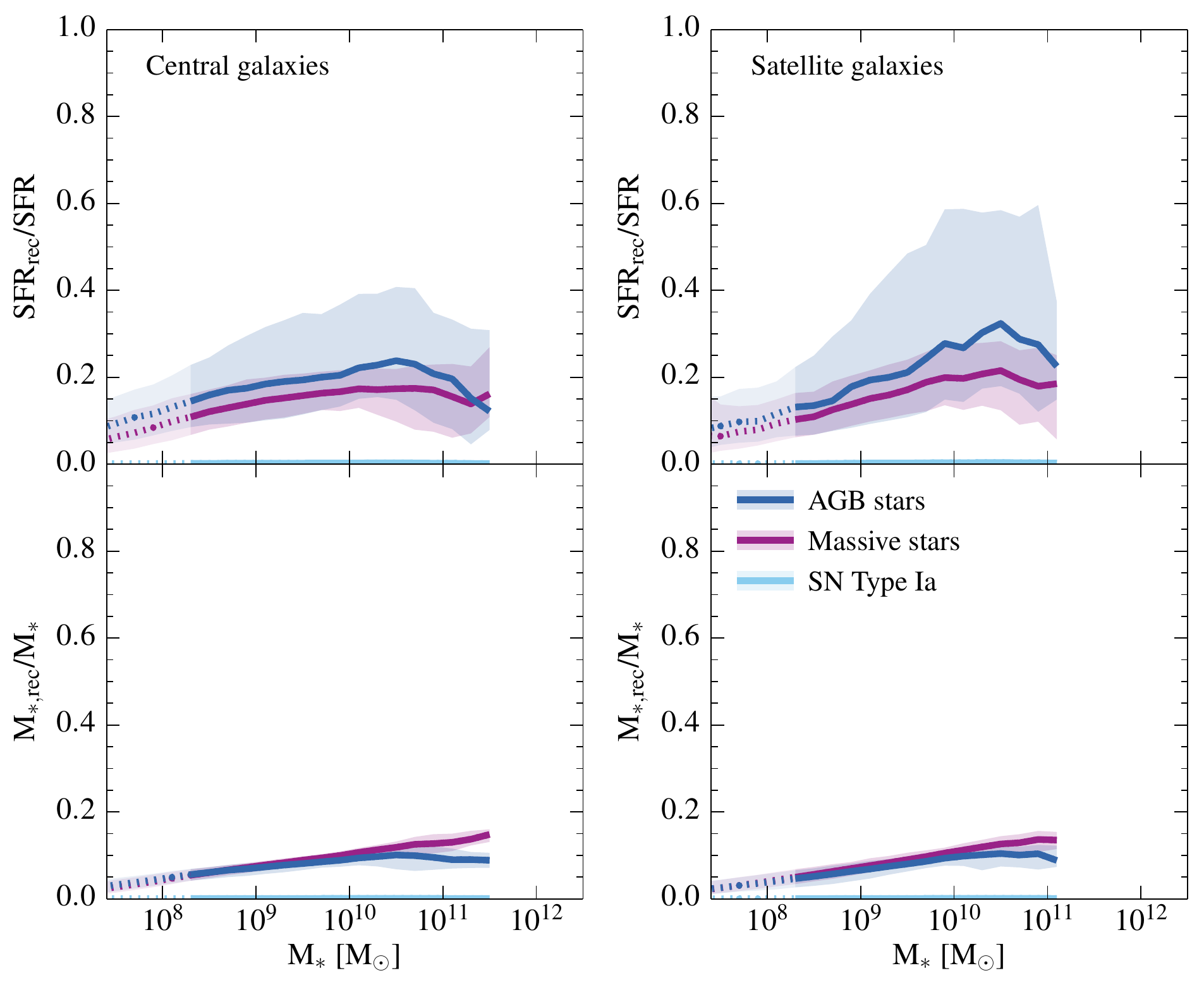}
\end{center}
\caption{The contribution of gas from AGB stars (blue), massive stars (purple) and SN Type Ia (cyan) to the SFR (top) and stellar mass (bottom) of galaxies at $z=0$ as a function of their stellar mass. The results for centrals and satellites are shown in the left and right panels, respectively. The curves and shaded regions indicate the medians and $10$th to $90$th percentile ranges, as in Fig.~\ref{fig:massdep_eagle}. In general, the gas from AGB stars and massive stars contributes about equally to the SFR and stelllar mass in both centrals and satellites. However, there is a slight enhancement in the contribution from AGB stars to the SFR at all but the highest mass scales, due to the preferential removal of massive star ejecta by star formation-driven winds and by lock up in stellar remnants. AGB ejecta are also responsible for the high \sfrratio values of some satellites, since these environmentally quenched objects have low sSFRs and cannot accrete gas. At the high-mass end, the relative contribution from AGB ejecta declines, because AGN feedback (which is unimportant at low mass) can drive them out even in the absence of star formation.}
\label{fig:massdep_AGB_SN}
\end{figure*}

To assess the relative significance of the different stellar mass loss channels for fuelling star formation in present-day centrals and satellites, we show in Fig.~\ref{fig:massdep_AGB_SN} the contribution of recycled gas to the SFR (top panels) and stellar mass (bottom panels) split into the contributions from AGB stars (blue), massive stars (purple) and SN Type Ia (cyan). These are plotted as a function of galaxy stellar mass for centrals (left panels) and satellites (right panels) at $z=0$.

From the top panels, we see that AGB stars are of greater importance for fuelling present-day star formation than massive stars in all but the most massive central galaxies. Up to $24 \%$ ($32 \%$) of the SFR in centrals (satellites) is fuelled by gas recycled from AGB stars, while $\lesssim 17 \%$ ($\lesssim 20 \%$) is fuelled by gas from massive stars. Integrated over cosmic history (as quantified by \mstarratio in the lower panels), their contributions are approximately equal at all but the highest mass scales. This may appear difficult to reconcile with the timed mass release from intermediate-mass and massive stars for a single SSP presented in Fig.~\ref{fig:SSP_chabrier}, where we showed that massive stars are the dominant source of (integrated) mass loss for SSP ages $\lesssim 1$ Gyr and that massive stars and AGB stars contribute about equally at higher ages. Fig.~\ref{fig:massdep_AGB_SN} implies that stellar ejecta do not simply accumulate in the ISM, but that they are removed, either by star formation or by outflows, in a way that affects the ejecta from AGB stars and massive stars differently\footnote{As our findings in Sections~\ref{sec:massdep_cen} and \ref{sec:massdep_sat} already imply, the relation between the mass loss from individual SSPs and the contribution of this mass loss to the SFR and stellar mass on galactic scales is not straightforward, since the rate of gas accretion and the efficiencies of stellar and AGN feedback depend on the mass of the galaxy.}.

The ejecta from massive stars are released almost instantaneously compared to those from AGB stars (see Fig.~\ref{fig:SSP_chabrier}). This means that, before AGB stars start to contribute significantly to the recycled gas, secondary generations of stars will already have formed from the massive star ejecta, causing an increasing fraction of these ejecta to become locked up in stellar remnants. While they then still contribute to the stellar mass of the galaxy, they no longer fuel ongoing star formation. Furthermore, a considerable amount of gas from massive stars will already have been ejected in galactic winds, before the AGB stars shed most of their mass. As a result, the contribution from massive stars to the present-day SFR is suppressed compared to that from AGB stars.

In general, if stars are older, the surrounding gas will contain a higher fraction of ejecta from intermediate-mass stars than from massive stars. This means that, at fixed stellar mass, galaxies with a low SFR (which includes passive galaxies) are expected to contain enhanced fractions of AGB ejecta, as only newly formed stars produce massive star ejecta, while evolved, intermediate-mass stars still shed mass in AGB winds. Not surprisingly, we see that the scatter to high values of \sfrratio for satellites with masses $10^{10}$ M$\sub{\odot} < M\sub{\ast} < 10^{11}$ M$\sub{\odot}$ (upper-right panel of Fig.~\ref{fig:massdep_eagle}) is mainly due to satellites with large fractions of the AGB ejecta being recycled, consistent with their low sSFRs (Fig.~\ref{fig:sat_massbin})\footnote{Despite the shallow, but significant, decrease in the sSFRs with mass of galaxies with $M\sub{\ast} \lesssim 10^{10.5}$ M$\sub{\odot}$ (shown in fig. 11 of S15), we do not see an increase in the \sfrratio from AGB stars \emph{relative} to \sfrratio from massive stars in this mass range. Note that we are now considering the \emph{ratio} of the blue and purple curves in the upper and lower left panels (focussing on central galaxies). Instead, the relative contribution from massive stars, both to the SFR (upper panels) and stellar mass (lower panels), remains approximately constant. This is due to a competing effect, namely the decrease in star formation feedback efficiency, which mitigates the preferential expulsion of massive star ejecta.}.

Finally, in massive galaxies ($M\sub{\ast} \gtrsim 10^{10.5}$ M$\sub{\odot}$), there is a decline of the \sfrratio and \mstarratio contributed by AGB stars, while the contributions from massive stars decrease only mildly (or flatten). This is consistent with the increase in the $[\mathrm{O}/\mathrm{Fe}]$ abundance ratio, as shown in Fig.~\ref{fig:ab_ratio} of Section~\ref{sec:metallicity}, where we discussed that this change in the relative significance of the different mass loss channels introduces a mass dependence in the relation between recycling-fuelled star formation and metallicity. We attribute this effect to `downsizing', a scenario in which the bulk of the stars in more massive galaxies has formed earlier and over a shorter period of time than in lower-mass counterparts \citep[e.g.][]{cowie_1996,neistein_2006,cattaneo_2008,fontanot_2009}.

The rapid and efficient star formation in the progenitors of present-day massive galaxies is however suppressed at later times, when these progenitors have grown massive enough for AGN feedback to become efficient and the gas cooling rates to drop. Moreover, while winds driven by feedback from star formation will not be available to drive out AGB ejecta in quenched galaxies, AGN feedback can. Hence, this scenario is consistent with the reduction in the contribution from AGB stars, relative to that from massive stars, to both the SFR (upper panels) and the stellar mass (lower panels) at the highest mass scales shown in Fig.~\ref{fig:massdep_AGB_SN}.

We infer that galaxies generally obtain most of their metals from massive star ejecta, as these ejecta have $4-6$ times higher metallicity than those from AGB stars, while the fraction of the (total) ISM mass and stellar mass contributed by massive star ejecta is similar to that contributed by AGB ejecta. This holds for metals in the gas-phase, but even more so for metals in the stellar phase. In both cases, the metal content contributed by massive star ejecta increases at the high-mass end ($M\sub{\ast} \gtrsim 10^{10.5}$ M$\sub{\odot}$) of the mass-metallicity relation. This is reflected by the trend of $[\mathrm{O}/\mathrm{Fe}]$ with stellar mass as presented in Fig.~\ref{fig:ab_ratio}, and is consistent with the abundance ratio trends observed for early-type galaxies \citep[e.g.][]{schiavon_2007,thomas_2010,johansson_2012,conroy_2014}. Our results imply that this $\alpha$-enhancement of massive galaxies is a consequence of AGN feedback.

\subsubsection{Radial dependence of gas recycling}
\label{sec:massdep_radius}

\begin{figure}
\begin{center}
\includegraphics[width=\columnwidth]{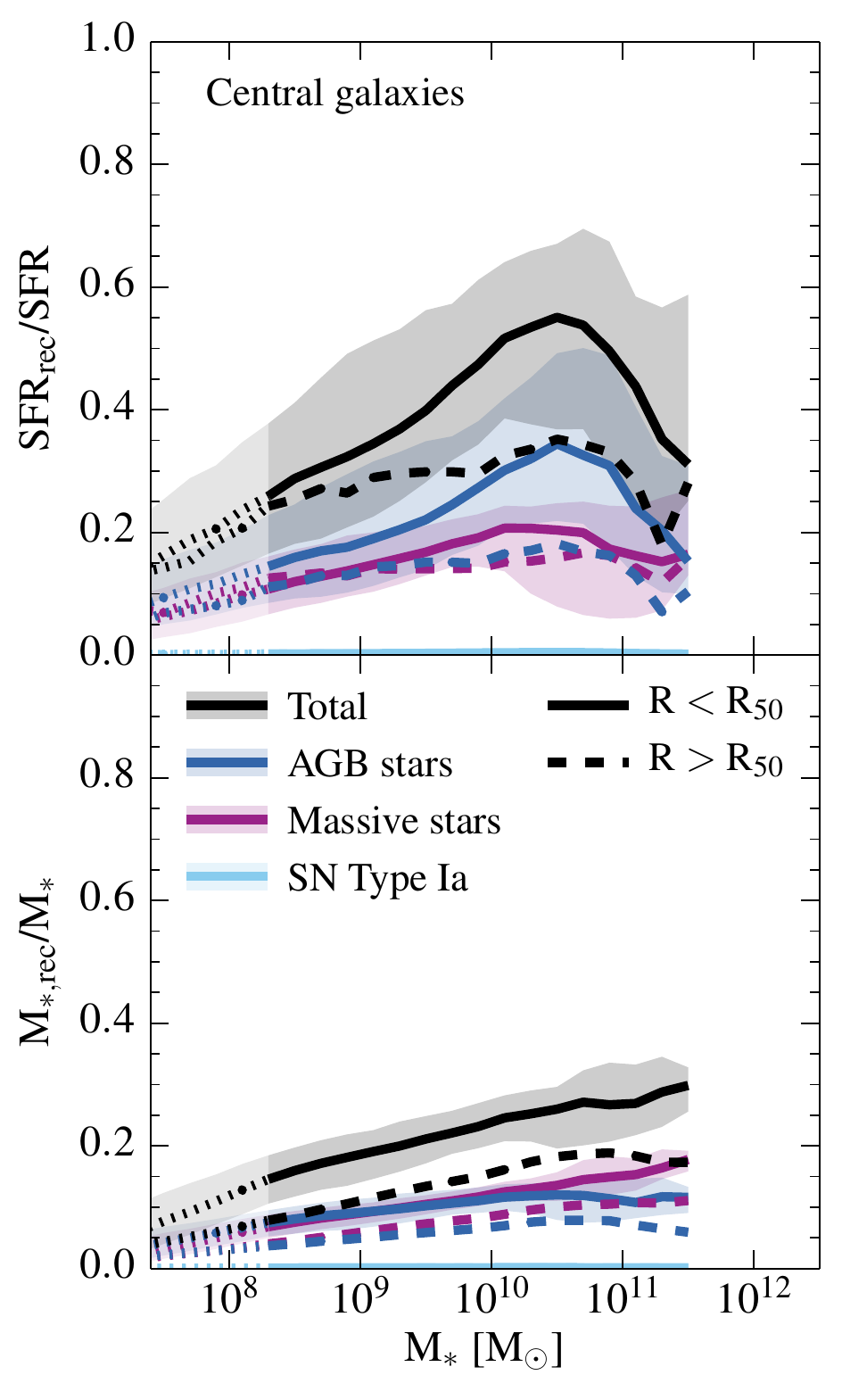}
\end{center}
\caption{As the left column of Fig.~\ref{fig:massdep_AGB_SN}, but split into gas and stars inside the stellar half-mass radius $R\sub{50}$ (solid) and outside $R\sub{50}$ (dashed). The $10$th to $90$th percentile ranges are only shown for inside $R\sub{50}$. Gas recycling is more important for fuelling star formation (at the present-day and in the past) in the inner parts of galaxies than in the outskirts. Consistent with inside-out growth, the gas in the central regions is comprised of an enhanced fraction of AGB ejecta, which is the main driver of the greater contribution of gas recycling to the star formation within $R\sub{50}$.}
\label{fig:massdep_radius}
\end{figure}

Having explored the importance of gas recycling on galaxy-wide scales, we now briefly investigate how the significance of recycling for fuelling star formation depends on the distance from the galactic centre. Note that the robustness of the results that we present in this section depends on the ability of the simulation to reproduce observed metallicity gradients. In addition, the results are subject to numerical uncertainties, like the mixing of metals, which may be underestimated by SPH simulations \citep[see][]{wiersma_2009b}. This will be investigated in a future paper. In Fig.~\ref{fig:massdep_radius} we plot, similar to the left column of Fig.~\ref{fig:massdep_AGB_SN}, the contribution of gas from AGB stars (blue), massive stars (purple) and SN Type Ia (cyan) to the SFR (top) and stellar mass (bottom) of centrals at $z=0$, now separated into gas and stars inside (solid lines) and outside (dashed lines) the stellar half-mass radius. This radius, denoted by $R\sub{50}$, is the 3D radius that encloses $50 \%$ of the stellar mass bound to the subhalo (within the $30$ pkpc 3D aperture). It is typically $\sim 4$ pkpc for a $M\sub{\ast} \sim 10^{10}$ M$\sub{\odot}$ galaxy. Note that we split both the numerator and the denominator of \sfrratio and \mstarratio into $R < R\sub{50}$ and $R > R\sub{50}$. We also plot the total recycled gas contribution (black), which is the sum over the three stellar mass loss channels, to the SFR and stellar mass in these two radial regimes.

Focusing first on the total (black lines), both panels consistently show that gas recycling is more important for fuelling star formation (at the present-day and in the past) in the central parts of galaxies than in the outskirts. This is consistent with the observational inference that galaxies grow in an inside-out fashion \citep[e.g.][]{munoz-mateos_2007,patel_2013}, with the oldest stars residing in the centre and the replenishment of the gas reservoir by late infall being primarily significant in the outskirts of the disc (owing to its relatively high angular momentum). From the black curves in the upper panel we see that, at the peak value, $55-60 \%$ of the SFR inside $R\sub{50}$ is due to stellar mass loss, compared to only $35-40 \%$ outside $R\sub{50}$. As a result, also a higher fraction of the stellar mass inside $R\sub{50}$ comes from recycling, $\sim 30 \%$ at the mass scale where recycling is most significant, compared to $\sim 20 \%$ in the outskirts.

To investigate what drives this radial dependence, we turn to the relative significance of the different sources of mass loss. While outside $R\sub{50}$ AGB stars (blue, dashed) and massive stars (purple, dashed) contribute about equally to the SFR for all masses, inside $R\sub{50}$ the contribution from AGB stars (blue, solid) is significantly larger than the contribution from massive stars (purple, solid). As discussed in the previous section, the gas around older stellar populations contains higher fractions of AGB ejecta. Hence, the difference between the inner and outer parts reflects the radial age gradient of the stars, due to the inside-out growth of the galaxy.

The drop in the AGB contribution at the high-mass end that we saw in Fig.~\ref{fig:massdep_AGB_SN}, is present in both the inner and outer parts, but it is much stronger near the galactic centre. This is consistent with it being due to a lack of intermediate-age stars and the ability of AGN to drive out the AGB ejecta even in the absence of star formation.

The radial variation of \sfrratio and \mstarratio is consistent with the negative metallicity gradients observed in local disc galaxies \citep[e.g.][]{zaritsky_1994,moustakas_2010,sanchez_2014}. Our results imply that, although the majority of the metals in galaxies (both in the ISM and stars) comes from massive star ejecta, the relative contribution from AGB ejecta is on average larger near the galactic centre than in the outskirts. This holds in particular for the ISM metal content of Milky Way-like ($M\sub{\ast} \sim 10^{10.5}$ M$\sub{\odot}$) galaxies.


\section{Exploring model variations with OWLS}
\label{sec:massdep_owls}

\begin{table*}
\centering
\caption{Set of OWLS simulations that vary in terms of the feedback implementation (upper section) or the adopted IMF (lower section). From left to right, the columns show the model name, whether or not there is energy feedback associated with star formation (SF feedback), metal-line cooling and AGN feedback, the adopted IMF, the fraction of kinetic energy available from SN Type II that is used to drive galactic winds ($f\sub{th}$), the initial wind velocity ($v\sub{w}$) and the wind mass loading parameter ($\eta$).}
\begin{tabular}{l r r r r r r r}
\hline
Name & SF feedback & Metal-line cooling & AGN feedback & IMF & $f\sub{th}$ & $v\sub{w}$ & $\eta$\\
& & & & & & [km/s] & \\
\hline
Feedback variations\\
\emph{REF} & $\checkmark$ & $\checkmark$ & $-$ & Chabrier & $0.40$ & $600$ & $2.0$\\
\emph{NOZCOOL} & $\checkmark$ & $-$ & $-$ & Chabrier & $0.40$ & $600$ & $2.0$\\
\emph{NOSN} & $-$ & $\checkmark$ & $-$ & Chabrier & $0.40$ & $600$ & $2.0$\\
\emph{NOSN\_NOZCOOL} & $-$ & $-$ & $-$ & Chabrier & $0.40$ & $600$ & $2.0$\\
\emph{AGN} & $\checkmark$ & $\checkmark$ & $\checkmark$ & Chabrier & $0.40$ & $600$ & $2.0$\\
\hline
IMF variations\\
\emph{IMFSALP} & $\checkmark$ & $\checkmark$ & $-$ & Salpeter & $0.66$ & $600$ & $2.0$\\
\emph{DBLIMFCONTSFML14} & $\checkmark$ & $\checkmark$ & $-$ & Top-heavy$^a$ & $0.40$ & $600$ & $14.6$\\
\emph{DBLIMFCONTSFV1618} & $\checkmark$ & $\checkmark$ & $-$ & Top-heavy$^a$ & $0.40$ & $1618$ & $2.0$\\
\hline
\end{tabular}
\label{tab:owls_models}

{$^a$\footnotesize{At high pressures ($P/k > 2.0 \times 10^6$ cm$^{-3}$ K) the IMF switches from \citet{chabrier_2003} to a top-heavy power-law $dN/dM \propto M^{-1}$.}}

\end{table*}

In this section we assess the sensitivity of our results from the EAGLE simulation presented in Section~\ref{sec:eagle} to the physical processes included in the subgrid model. We do this by comparing a set of OWLS simulations \citep{schaye_2010} in which the subgrid model is systematically varied. In particular, we explore variations of the feedback from star formation and AGN (Section~\ref{sec:massdep_feedback}), in combination with metal-line cooling, by explicitly turning on or off a particular process. The variation of \sfrratio and \mstarratio with galaxy mass in Figs.~\ref{fig:massdep_eagle_reso}, \ref{fig:massdep_eagle}, \ref{fig:massdep_AGB_SN} and \ref{fig:massdep_radius} is already suggestive of the important role played by these feedback processes and we will now show this explicitly. We also vary the adopted IMF (Section~\ref{sec:massdep_imf}), as this determines the fraction of the stellar mass that is released by a stellar population\footnote{We use the OWLS simulations to perform the model comparison, as this suite provides both the extreme feedback variations and IMF variations we need. Note that while the stellar and AGN feedback models in OWLS are slightly different from those in EAGLE, and that OWLS does not reproduce the observed $z \simeq 0$ GSMF, we can still use the OWLS suite to study the \emph{relative} changes in \sfrratio and \mstarratio (at least qualitatively).}. The sets of variations are summarized in Table~\ref{tab:owls_models} and are described in more detail below.

We start this section with a brief overview of the OWLS simulation set-up and the implemented subgrid physics, focusing in particular on the differences with respect to EAGLE (for a detailed description of the differences, we refer the reader to S15). The OWLS simulations were run with a modified version of the SPH code \textsc{Gadget3}, but in contrast to EAGLE it uses the entropy formulation of SPH implemented by \citet{springel_hernquist_2002}. The adopted cosmological parameters, $\left[ \Omega\sub{m},\Omega\sub{b},\Omega\sub{\Lambda},\sigma\sub{8},n\sub{s},h \right]=\left[ 0.238,0.0418,0.762,0.74,0.951,0.73 \right]$, are consistent with WMAP 3-year \citep{spergel_2007} and WMAP 7-year data \citep{komatsu_2011}. The simulations used here were run in periodic volumes of size $L = 100$ $h^{-1}$ cMpc.\footnote{Note that for OWLS, the box size and particle masses are given in units with $h^{-1}$.}, containing $N = 512^3$ dark matter particles with initial mass $m\sub{dm}=4.1 \times 10^8\ h^{-1}\ {\rm M}\sub{\odot}$ and an equal number of baryonic particles with initial mass $m\sub{b}=8.7 \times 10^7\ h^{-1}\ {\rm M}\sub{\odot}$. The gravitational softening length is $7.81$ $h^{-1}$ ckpc, limited to a maximum of $2.00$ $h^{-1}$ pkpc.

The implementations of radiative cooling and heating, and stellar evolution, are nearly the same as in EAGLE. Hence, the timed mass release by an SSP, as presented in Section~\ref{sec:ssp}, is nearly identical in OWLS. On the other hand, differences in the OWLS subgrid physics include the use of a fixed density threshold in the implementation of star formation and the kinetic implementation of star formation-driven winds. A fixed fraction $f\sub{th}$ of the available feedback energy is injected locally, where it directly (by `kicking' gas particles surrounding newly-formed star particles) generates galactic winds with initial velocity $v\sub{w}$ and mass loading $\eta$ \citep[following][]{dallavecchia+schaye_2008}. The prescriptions for BH growth and AGN feedback \citep{booth+schaye_2009}, which are only included in the OWLS model \emph{AGN}, employ a Bondi-Hoyle accretion rate with a density-dependent correction term, and a slightly lower thermal heating temperature than in EAGLE.


\subsection{Effect of feedback processes and metal-line cooling}
\label{sec:massdep_feedback}

We consider the following set of feedback variations, combined with variations in the metal-line cooling, as the latter may also impact upon the efficiency of the feedback:

\begin{itemize}
\item \emph{REF} is the OWLS fiducial model, which serves as the reference in the model comparison. It includes radiative cooling and heating, star formation, stellar evolution and kinetic energy feedback (with $f\sub{th} = 0.40$, $v\sub{w}=600$ km s$^{-1}$ and $\eta=2$) from star formation. Note that it does \emph{not} include prescriptions for BH growth and AGN feedback. All model variations listed below are varied with respect to this model.
\item In \emph{NOZCOOL} the cooling rates are calculated using primordial element abundances, i.e. $X = 0.76$ and $Y = 0.24$.
\item \emph{NOSN} turns off all energy feedback mechanisms associated with star formation. However, the mass loss and metal production by massive stars, SN Type Ia and AGB stars are still present.
\item \emph{NOSN\_NOZCOOL} is a combination of the previous two models. It does not include galactic winds and the cooling rates are based on primordial element abundances.
\item \emph{AGN} includes models for the growth of BHs and feedback from AGN.
\end{itemize}

\begin{figure}
\begin{center}
\includegraphics[width=\columnwidth]{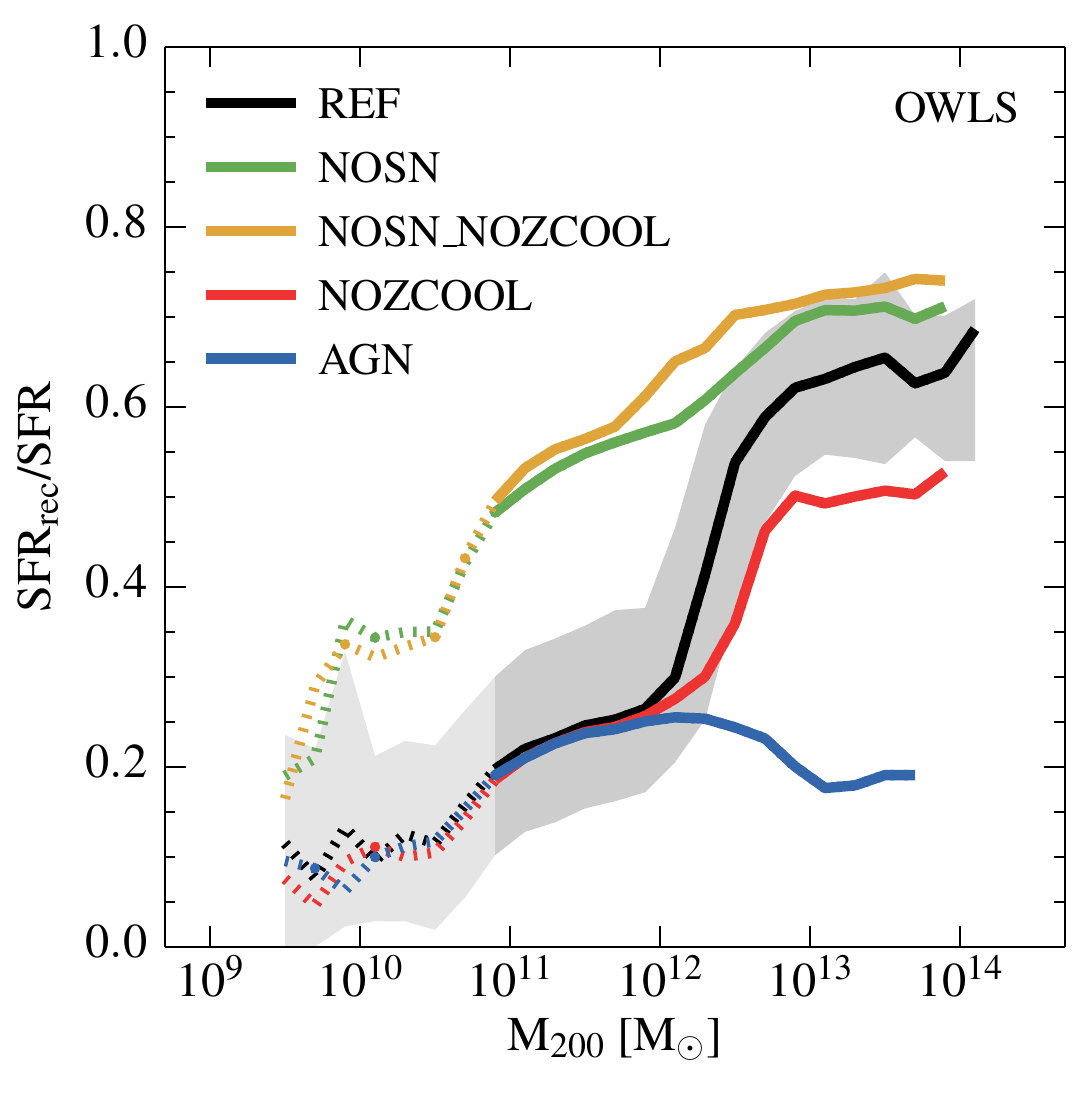}
\end{center}
\caption{Comparison of a set of OWLS models to explore the effects of star formation feedback, AGN feedback and metal-line cooling on the contribution of recycled gas to the SFR at $z=0$. The curves show the median (in logarithmic mass bins of size $0.2$ dex containing at least $10$ haloes) contribution of recycled gas to the SFR as a function of halo mass. The grey shaded region shows the $10$th to $90$th percentile region for the OWLS fiducial model (black curve). The solid curves become dotted when the halo mass corresponds to fewer than $100$ dark matter particles. We find that the efficiency of the feedback associated with star formation (in low-mass galaxies) and the feedback of AGN (in high-mass galaxies) determines how much gas from stellar mass loss contributes to the SFR. A higher feedback efficiency results in a lower contribution from recycled gas.}
\label{fig:massdep_feedback}
\end{figure}

Fig.~\ref{fig:massdep_feedback} shows the effects of star formation feedback, AGN feedback and metal-line cooling on \sfrratio as a function of halo mass, $M\sub{200}$\footnote{Note that we do not treat central and satellite galaxies separately, but instead consider the total amount of star formation taking place within the group halo as a whole (without applying an aperture when calculating \sfrratio). Also, while we only present the results for \sfrratio as a function of halo mass, they are consistent with the results for \mstarratio, as well as for both ratios as functions of stellar mass.}. Note that the mass scale corresponding to $100$ dark matter particles, below which the curves are shown as dotted lines, is a factor of $\sim 58$ greater than in the EAGLE fiducial simulation.

First comparing \emph{NOSN} (green curve) to \emph{REF} (black curve), we see that the feedback associated with star formation dramatically reduces the contribution of recycled gas to the SFR at all mass scales, in particular for $M\sub{200} \lesssim 10^{12}$ M$\sub{\odot}$, where the regulation of star formation is largely governed by feedback from star formation. Here, the contribution drops from $\sim 55 \%$ (without star formation feedback, \emph{NOSN}) to $\sim 25 \%$ (with star formation feedback, \emph{REF}). The large reduction can be explained if we consider the environments from which these star formation-driven winds are launched. The outflows originate in the dense ISM, which is the environment into which stellar mass loss is deposited. These stellar ejecta, which could be recycled into new generations of star formation, are prevented from forming stars if the winds eject the gas from the ISM in such a way that it does not return and become sufficiently dense again on short timescales. If, on the other hand, star formation feedback is inefficient, then this gas remains in the ISM and fuels star formation. The \emph{REF} model shows an increasing \sfrratio with mass, which becomes increasingly similar to the \emph{NOSN} model. This is consistent with SN feedback becoming less efficient as the depth of the potential well and the density and pressure in the ISM increase. Hence, the decreasing efficiency of the feedback leads to gas recycling being increasingly important for fuelling star formation in more massive systems.

A notable feature in the curve of the \emph{REF} model is the sharp upturn of \sfrratio at a halo mass of $M\sub{200} \sim 10^{12}$ M$\sub{\odot}$. Clearly, the feedback suddenly becomes very inefficient at this mass scale. As explained by \citet{dallavecchia+schaye_2012}, this is due to strong artificial radiative losses in the ISM as the gas (which has a high pressure and density in these high-mass systems) gets shock-heated by the star formation-driven winds. Kinetic energy is thermalised to temperatures at which the cooling time is short relative to the sound crossing time, and is quickly radiated away. As a consequence, the winds stall in the ISM before they can escape the galaxy, which drives up the value of \sfrratio. For a fixed initial velocity of the kinetically implemented winds, this effect causes a sharp transition at the mass scale for which artificial losses become significant. This results in unrealistic stellar mass fractions in haloes of $M\sub{200} > 10^{12}$ M$\sub{\odot}$ \citep{haas_2013} and a failure of the model to reproduce the observed GSMF \citep{crain_2009}. Comparing \emph{REF} to \emph{NOZCOOL} (red), we see that the upturn becomes less pronounced if the cooling rates are reduced. Lower cooling rates, as a result of neglecting metal-line cooling (\emph{NOZCOOL}), reduce the artificial thermal losses in the ISM and therefore enable the feedback to remain more efficient to higher mass scales. This results in lower values of \sfrratio in the most massive systems.

In addition to its impact upon the feedback efficiency, a change in the cooling rates also affects the accretion rate onto the galaxy, therefore impacting upon \sfrratio in a more direct fashion. However, from comparing \emph{NOSN} (with metal-line cooling; green) and \emph{NOSN\_NOZCOOL} (without metal-line cooling; yellow) we see that, in the absence of energy feedback associated with star formation, the effect of changing the cooling rates on \sfrratio is small, especially considering the expected amount of scatter in the two relations (from the grey shaded region). Hence, we conclude that a change in the cooling rates, as a result of turning metal-line cooling on or off, mainly affects the contribution of recycled gas to the SFR by changing the (partly numerical) efficiency of the star formation feedback implementation.

Finally, to investigate the effect of AGN feedback on \sfrratio as a function of halo mass, we compare the \emph{AGN} model (blue curve), which includes AGN feedback, to the \emph{REF} model. Since BHs live in the dense, central regions of galaxies, where a large fraction of the stellar ejecta are deposited, we again expect a low recycling-fuelled SFR if the AGN feedback is strong enough to eject gas from the ISM. The \emph{AGN} model curve indeed shows that at masses $M\sub{200} \gtrsim 10^{12}$ M$\sub{\odot}$, where feedback from AGN becomes important, the values of \sfrratio decrease towards higher masses. At a halo mass of $M\sub{200} \sim 10^{13}$ M$\sub{\odot}$, recycled gas contributes only $\sim 20 \%$ to the SFR. This is in stark contrast to the \emph{REF} model, for which the contribution reaches $\sim 65 \%$, indicating the strong impact of AGN feedback on the \sfrratio in the regime where the feedback from star formation is inefficient and AGN are the main drivers of galactic winds. This highlights the importance of including AGN feedback in the subgrid model. Qualitatively, we conclude that for massive galaxies the increasing efficiency of AGN feedback towards higher masses leads to gas recycling being less important for fuelling present-day star formation.


\subsection{Effect of changing the stellar initial mass function}
\label{sec:massdep_imf}

We consider the following variations with respect to the fiducial \citet{chabrier_2003} IMF:

\begin{itemize}
\item \emph{IMFSALP} adopts a \citet{salpeter_1955} IMF, spanning the same stellar mass range as the Chabrier IMF used by the fiducial model. The corresponding change in the amplitude of the observed \citet{kennicutt_1998} relation is taken into account. While $v\sub{w}$ and $\eta$ in the implementation of star formation-driven winds are kept the same, as is the total wind energy per unit stellar mass (which is proportional to $v\sub{w}\eta\up{2}$), $f\sub{th}$ is increased to $0.66$.
\item \emph{DBLIMFCONTSFML14} assumes an IMF that becomes top-heavy in high-pressure environments. For $P/k > 2.0 \times 10^6$ cm$^{-3}$ K the IMF switches from a Chabrier IMF to a power-law $dN/dM \propto M^{-1}$. In these environments there is $7.3$ times more stellar feedback energy available per unit stellar mass to drive galactic winds. In this model the additional energy is used to increase the wind mass loading by a factor of $7.3$.
\item As the previous model, \emph{DBLIMFCONTSFV1618} switches to a top-heavy power-law IMF for stars forming in high-pressure regions. However, the additional stellar feedback energy is now used to increase the initial velocity of the winds: $v\sub{w}$ is a factor of $\sqrt{7.3}$ higher than in the reference model.
\end{itemize}

\begin{figure}
\begin{center}
\includegraphics[width=\columnwidth]{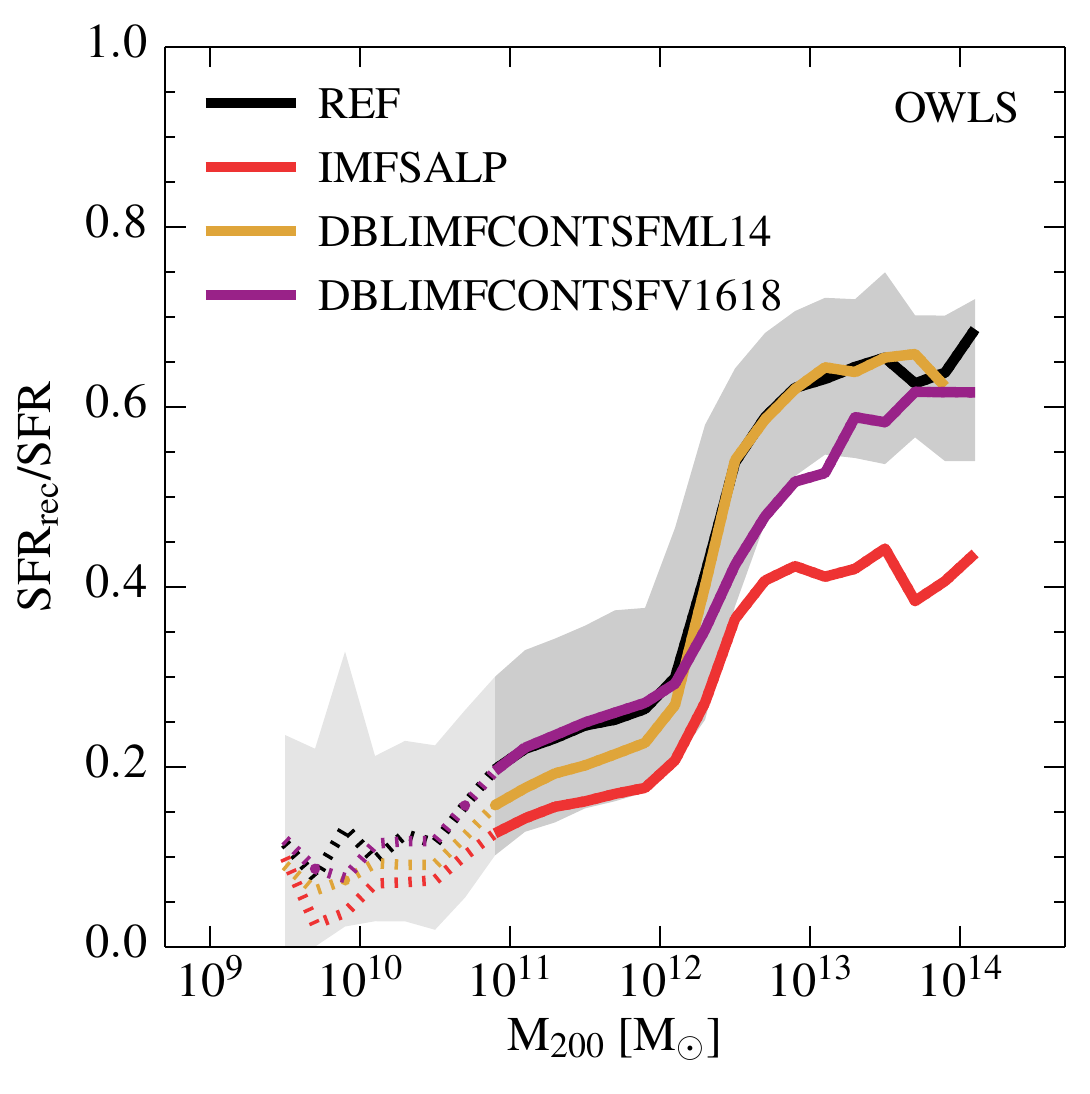}
\end{center}
\caption{As Fig.~\ref{fig:massdep_feedback}, but showing a set of OWLS models with different IMFs. We find that adopting a more bottom-heavy (top-heavy) IMF reduces (enhances) the contribution from recycled gas to the SFR, but only if the feedback energy used to initiate star formation-driven galactic winds is kept fixed. If the extra stellar feedback energy from adopting a top-heavy IMF is used to either increase the mass loading or the wind velocity, then the recycled gas contributions decrease, showing that other IMF-related effects are more than compensated for by the increased efficiency of the star formation feedback.}
\label{fig:massdep_imf}
\end{figure}

Fig.~\ref{fig:massdep_imf} shows the effect of varying the choice of IMF. A comparison of \emph{IMFSALP} (red curve) and \emph{REF} (black curve) shows that adopting a more bottom-heavy IMF like Salpeter reduces the values of \sfrratio over the whole mass range. This was expected, because both the total mass and the metal mass released by stellar populations are lower than for a Chabrier IMF (compare Figs.~\ref{fig:SSP_chabrier} and \ref{fig:SSP_salpeter}). In low-mass galaxies (in haloes with masses $M\sub{200} \lesssim 10^{12}$ M$\sub{\odot}$) the contribution of recycled gas to the SFR drops from $\sim 25 \%$ to $\sim 15 \%$, while in high-mass galaxies it drops from $\sim 65 \%$ to $\sim 40 \%$. In the high-mass systems, we expect that the reduction is partly due to a reduction of the cooling rates, which causes a slight increase of the efficiency of star formation feedback. This is the result of the reduced metal mass released by stellar populations, similar to disabling metal-line cooling in the \emph{NOZCOOL} model (see Section~\ref{sec:massdep_feedback}). However, note that since the total wind energy per unit stellar mass is kept fixed when switching from \emph{REF} to the \emph{IMFSALP} model, the feedback efficiency is affected only by a change in the cooling rates.

On the other hand, the switch to a top-heavy IMF as implemented in \emph{DBLIMFCONTSFML14} and \emph{DBLIMFCONTSFV1618}, leads to competing effects. A top-heavy IMF yields more stellar mass loss, but also yields more stellar feedback energy to drive galactic winds, which is partly dissipated due the increased metal mass loss. Comparing \emph{REF} with \emph{DBLIMFCONTSFML14} (increased $\eta$) and \emph{DBLIMFCONTSFV1618} (increased $v\sub{w}$) enables us to determine which effect dominates. We have investigated (but do not show here) the effects of increasing the mass loading or the initial wind velocity without changing the IMF. We find that if the winds efficiently escape the galaxy (in low-mass systems, $M\sub{200} \lesssim 10^{12}$ M$\sub{\odot}$) the mass loading mainly determines the gas mass that is ejected, whereas the initial velocity is of little importance. On the other hand, in high-mass systems ($M\sub{200} \gtrsim 10^{12}$ M$\sub{\odot}$), where the artificial radiative losses are high and star formation-driven winds are not efficient in escaping the galaxy, we find that increasing the mass loading has little effect, as these losses remain too significant. Boosting instead the initial wind velocity, alleviates these losses and increases the efficiency of the wind, because the wind now thermalises at a higher temperature \citep{haas_2013}.

This is consistent with the IMF variations, \emph{DBLIMFCONTSFML14} (yellow curve) and \emph{DBLIMFCONTSFV1618} (purple curve), shown in Fig.~\ref{fig:massdep_imf}. For \emph{DBLIMFCONTSFML14} the values of \sfrratio are reduced at low masses and consistent with \emph{REF} at high masses, whereas for \emph{DBLIMFCONTSFML14} \sfrratio is consistent at low masses and reduced at high masses. Hence, we infer that the feedback efficiency is the dominant factor in determining \sfrratio as a function of halo mass. Despite the fact that, naively, we might have expected an \emph{increase} of \sfrratio upon adopting a top-heavy IMF in high-pressure regions, as a result of the larger fraction of the stellar material available for recycling, the \emph{decrease} of \sfrratio in the \emph{DBLIMFCONTSFML14} and \emph{DBLIMFCONTSFV1618} models compared to the \emph{REF} model shows that this effect is more than compensated for by the increased efficiency of the stellar feedback.

From the OWLS model comparisons presented in this section we conclude that the efficiency of the feedback from star formation and AGN is key for regulating the fuelling of star formation with recycled gas. The choice of the IMF sets the total mass of gas that is potentially available for recycling, but the feedback efficiency determines how much gas recycled from stellar mass loss actually contributes to the SFR (and hence the stellar mass) of galaxies. This makes the contribution from recycled gas to the SFR and stellar mass sensitive to the mass of the galaxy.


\section{Summary and discussion}
\label{sec:summary}

We have investigated the significance of stellar ejecta as fuel for star formation using the \emph{Ref-L100N1504} cosmological simulation from the EAGLE project. We studied the contribution of gas from evolved stellar populations to the SFR and stellar mass, as a cosmic average as a function of redshift and within individual galaxies as a function of metallicity and galaxy stellar mass at $z=0$. We treated the galaxies identified as `centrals' separately from those identified as `satellites'. Since the mass released by AGB stars, SN Type Ia and massive stars was explicitly followed in the simulation, we were able to assess the relative significance of these different mass loss channels for fuelling star formation. We also explored the radial dependence of gas recycling, by comparing the significance of recycling-fuelled star formation in the inner and outer parts of galaxies. Our results can be summarized as follows:

\begin{itemize}
\item The contribution of recycled gas to the present-day SFR and stellar mass of galaxies is strongly, positively correlated with, respectively, the metallicity of the ISM and stars. Therefore, many of our conclusions on the role of stellar ejecta in fuelling star formation as a function of galaxy mass and type carry over to the mass-metallicity relation. The relations between the contribution of stellar mass loss and metallicity do exhibit a slight dependence on galaxy stellar mass, as a result of the increasing contribution of mass loss from massive stars relative to that from intermediate-mass stars to the SFR and stellar mass for $M\sub{\ast} \gtrsim 10^{10.5}$ M$\sub{\odot}$ (Fig.~\ref{fig:massdep_AGB_SN}, Section~\ref{sec:massdep_AGB_SN}). We provide the best-fit relations (equations~\ref{eq:sfr_metal} and \ref{eq:mstar_metal}), including a term with the $[\mathrm{O}/\mathrm{Fe}]$ abundance ratio, which enable one to estimate the importance of gas recycling in present-day galaxies from the observed metallicity and $\alpha$-enhancement (Fig.~\ref{fig:metallicity}, Section~\ref{sec:metallicity}).
\item We apply the relations between the SFR contributed by recycling and ISM metallicity and between the stellar mass contributed by recycling and stellar metallicity from EAGLE to the observed mass-metallicity relations to estimate the recycled gas contributions as a function of galaxy stellar mass. Since we find these relations to be insensitive to the subgrid models for feedback, applying them to the observed mass-metallicity relations yields more accurate estimates for the contribution of recycling than the direct predictions of EAGLE, provided that the (systematic) uncertainty in the calibration of the observed mass-metallicity relation is smaller than the discrepancy between the mass-metallicity relation predicted by EAGLE and the observed relation. For central galaxies with a stellar mass similar to that of the Milky Way ($M\sub{\ast} \sim 10^{10.5}$ M$\sub{\odot}$), which corresponds to the mass scale of the peak in the galaxy formation efficiency, $35 \%$ of the present-day SFR and $20 \%$ of the present-day stellar mass is due to recycled stellar mass loss (Fig.~\ref{fig:massdep_eagle_reso}, Section~\ref{sec:massdep_cen}).
\item Recycling of stellar mass loss becomes increasingly important for fuelling star formation towards lower redshift. At the present-day, the fiducial EAGLE model (i.e. as computed directly from the simulation) indicates that approximately $35 \%$ of the cosmic SFR density and $19 \%$ of the cosmic stellar mass density is contributed by recycling (Figs.~\ref{fig:evo_sfr_mstar} and \ref{fig:evo_ratios}, Section~\ref{sec:evolution}).
\item The fraction of the present-day SFR and stellar mass of central galaxies contributed by recycling shows a characteristic trend with the mass of the galaxy and its subhalo: for $M\sub{\ast} \lesssim 10^{10.5}$ M$\sub{\odot}$ ($M\sub{sub} \lesssim 10^{12.2}$ M$\sub{\odot}$) the contribution increases with mass, while for $M\sub{\ast} \gtrsim 10^{10.5}$ M$\sub{\odot}$ the contribution turns over and decreases with mass (in case of the SFR) or remains approximately constant (in case of the stellar mass). We infer that this trend is regulated by the efficiency of the feedback associated with star formation (at low mass scales) and AGN (at high mass scales). If feedback is efficient in driving galactic winds and thereby ejecting gas from the ISM, which is the environment into which stellar mass loss is deposited, then this will preferentially reduce the SFR and stellar mass contributed by recycled gas (Figs.~\ref{fig:massdep_eagle_reso} and \ref{fig:massdep_eagle}, Section~\ref{sec:massdep_eagle}).
\item The importance of gas recycling for fuelling ongoing star formation in satellite galaxies is broadly consistent with that for central galaxies over a wide range of masses, as recycling is mainly governed by the efficiency of feedback. However, the fiducial EAGLE model indicates that in satellites with a Milky Way-like mass the fraction of the SFR contributed by recycled gas significantly exceeds the one in similarly massive centrals, and even reaches $\gtrsim 90 \%$ for satellites with the lowest gas fractions (Fig.~\ref{fig:massdep_eagle}, Section~\ref{sec:massdep_sat}). We infer that this results from a depletion of the ISM gas reservoir of the satellite, either due to the cessation of fresh infall or the removal of gas from the disc, which makes them more reliant on stellar mass loss for fuelling ongoing star formation (Fig.~\ref{fig:sat_massbin}, Section~\ref{sec:massdep_sat}).
\item As a cosmic average, the gas from AGB stars accounts for an increasing fraction of the recycled stellar mass loss towards lower redshift. At $z \gtrsim 0.4$, however, massive stars still provide the majority of the gas that fuels the cosmic SFR density through recycling. As a result, massive stars dominate the recycling-fuelled stellar mass density at all redshifts. The contribution from SN Type Ia is always small (Figs.~\ref{fig:evo_sfr_mstar} and \ref{fig:evo_ratios}, Section~\ref{sec:evolution}).
\item Within individual galaxies, AGB stars and massive stars contribute approximately equally to the present-day SFR and stellar mass of centrals and satellites. The contribution from AGB stars to the SFR is slightly enhanced with respect to the massive star contribution at all but the highest mass scales, which results from the preferential ejection of massive star ejecta by star formation-driven winds and their early lock up in stellar remnants. At the highest mass scales ($M\sub{\ast} \gtrsim 10^{10.5}$ M$\sub{\odot}$), on the other hand, we find a relative enhancement in the contributions from massive stars, which we attribute to a downsizing effect, with more massive galaxies forming their stars earlier and more rapidly. Their stellar mass therefore preferentially consists of massive star ejecta, which are recycled on short timescales (Fig.~\ref{fig:massdep_AGB_SN}, Section~\ref{sec:massdep_AGB_SN}).
\item Exploring the radial dependence of gas recycling within central and satellite galaxies, we find that recycling is more important for fuelling star formation (at the present-day and in the past) in the central parts of galaxies (within $R\sub{50}$) than in the outskirts (outside $R\sub{50}$), which is consistent with the observationally inferred inside-out growth of galaxies. We find that the difference between these two radial regimes is predominantly driven by the difference in the fractional contribution from AGB stars to the SFR (stellar mass), which is significantly higher than (roughly equal to) the one from massive stars inside $R\sub{50}$ and roughly equal (lower) outside $R\sub{50}$. This radial trend directly reflects the negative stellar age gradient with increasing distance from the galactic centre (Fig.~\ref{fig:massdep_radius}, Section~\ref{sec:massdep_radius}).
\end{itemize}

Finally, we assessed the sensitivity of our results from the EAGLE simulation to the physical processes in the subgrid model using a suite of simulations from the OWLS project. The suite is comprised of a set of extreme variations of the feedback model, in which star formation feedback, AGN feedback and metal-line cooling are switched on or off entirely (Fig.~\ref{fig:massdep_feedback}, Section~\ref{sec:massdep_owls}), as well as a set of variations of the adopted IMF (Fig.~\ref{fig:massdep_imf}, Section~\ref{sec:massdep_owls}). A systematic comparison of the results shows that while the total fraction of the stellar mass that is available for recycling is determined by the adopted IMF, the fraction of the stellar mass loss that is actually used to fuel star formation is controlled by the efficiency of the feedback associated with star formation and the feedback from AGN, each affecting the galaxy mass regime where the respective feedback process regulates the star formation.

Consistent with previous studies \citep[e.g.][]{kennicutt_1994,leitner+kravtsov_2011,voit+donahue_2011}, our results emphasize the importance of modelling the recycling of stellar ejecta in simulations of galaxy formation, and the necessity of accounting for such gas in assessments of the `fuel budget' of present-day galaxies. The fractional contribution of recycled ejecta to the SFR and stellar mass is not dominant, but it is also not negligible, and it extends the gas consumption timescale significantly beyond that implied by the ratio of the instantaneous gas mass and star formation rate of galaxies.

The relatively small contribution of recycling to the SFR and stellar mass of massive galaxies in our simulations is contrary to the naive expectation that the establishment of a hot circumgalactic medium quenches gas infall and renders the galaxy reliant on recycling for continued fuelling. Instead, the simulations indicate that the ongoing star formation in massive galaxies is sustained mostly by unprocessed gas. An interesting route for future studies will be to explore whether this gas originates in cooling flows, or is stripped from infalling satellite galaxies.


\section*{Acknowledgements}

We thank the anonymous referee for helpful comments. This work used the DiRAC Data Centric system at Durham University, operated by the Institute for Computational Cosmology on behalf of the STFC DiRAC HPC Facility (www.dirac.ac.uk). This equipment was funded by BIS National E-infrastructure capital grant ST/K00042X/1, STFC capital grant ST/H008519/1, and STFC DiRAC Operations grant ST/K003267/1 and Durham University. DiRAC is part of the National E-Infrastructure. We also gratefully acknowledge PRACE for awarding us access to the resource Curie based in France at Tr\'{e}s Grand Centre de Calcul. This work was sponsored with financial support from the Netherlands Organization for Scientific Research (NWO), from the European Research Council under the European Union's Seventh Framework Programme (FP7/2007-2013) / ERC Grant agreement 278594-GasAroundGalaxies, from the National Science Foundation under Grant No. NSF PHY11-25915, from the UK Science and Technology Facilities Council (grant numbers ST/F001166/1 and ST/I000976/1) and from the Interuniversity Attraction Poles Programme initiated by the Belgian Science Policy Office ([AP P7/08 CHARM]). RAC is a Royal Society University Research Fellow.


\bibliographystyle{mnras}
\bibliography{bibliography}


\appendix

\section{Mass released by an SSP with a Salpeter IMF}
\label{sec:ssp_salpeter}

\begin{figure*}
\begin{center}
\includegraphics[width=0.95\textwidth]{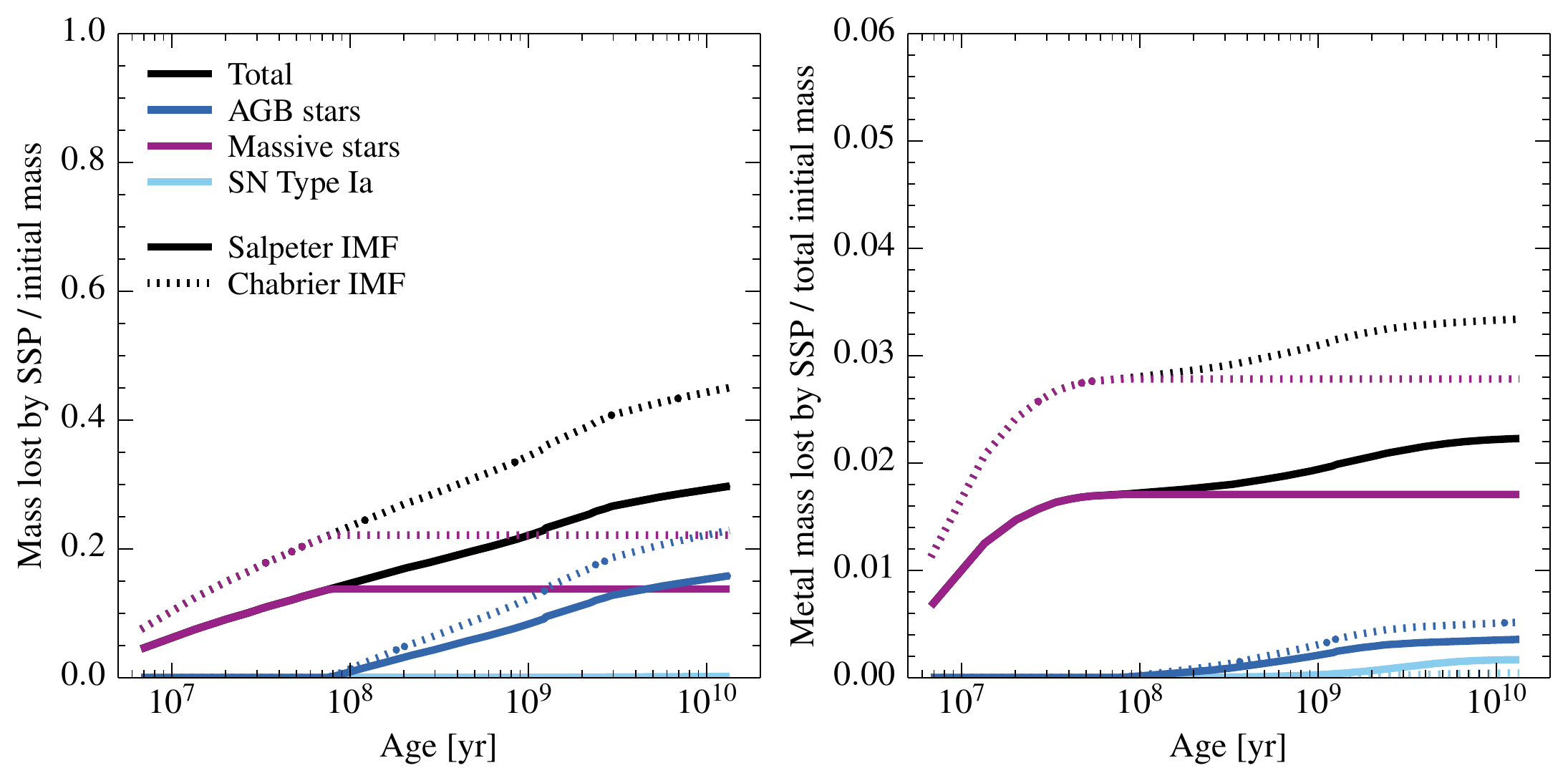}
\end{center}
\caption{The cumulative fraction of the initial mass (total: left panel; in the form of metals: right panel) that is released by an SSP as a function of its age, adopting a \citet{salpeter_1955} IMF (solid lines) or a \citet{chabrier_2003} IMF (dotted lines). The curves show the contributions from AGB stars (blue), massive stars (purple) and SN Type Ia (cyan), as well as the total (metal) mass ejected by the SSP (black) for solar stellar metallicity. The total (metal) mass loss is lower for a Salpeter IMF than for a Chabrier IMF by a factor of $\sim 1.5$. Adopting a Salpeter IMF increases the relative contributions from AGB stars and SN Type Ia, which have intermediate-mass progenitor stars.}
\label{fig:SSP_salpeter}
\end{figure*}

Fig.~\ref{fig:SSP_salpeter} shows the total (left) and metal (right) mass released by an SSP with a \citet{salpeter_1955} IMF (solid lines) and by an SSP with a \citet{chabrier_2003} IMF (dotted lines) in the range $0.1-100$ M$\sub{\odot}$ as a function of age for solar metallicity. The colours in both panels have the same meaning as in Fig.~\ref{fig:SSP_chabrier}. The (metal) mass loss is lower for a Salpeter IMF than for a Chabrier IMF by a factor of $\sim 1.5$ over the whole range of SSP ages plotted. The relative contributions from massive stars and AGB stars are somewhat lower and higher, respectively. The metal mass loss from SN Type Ia is higher for a Salpeter IMF, accounting for $\sim 8 \%$ of the total metal mass released. In general, adopting a Salpeter IMF instead of a Chabrier IMF increases the relative contributions from intermediate-mass stars to the (metal) mass loss, as expected for a more bottom-heavy IMF.


\section{Numerical convergence}
\label{sec:resotests}

\subsection{Relation between recycled gas contributions and stellar mass}
\label{sec:resotests_massdep}

\begin{figure*}
\begin{center}
\includegraphics[width=0.9123\textwidth]{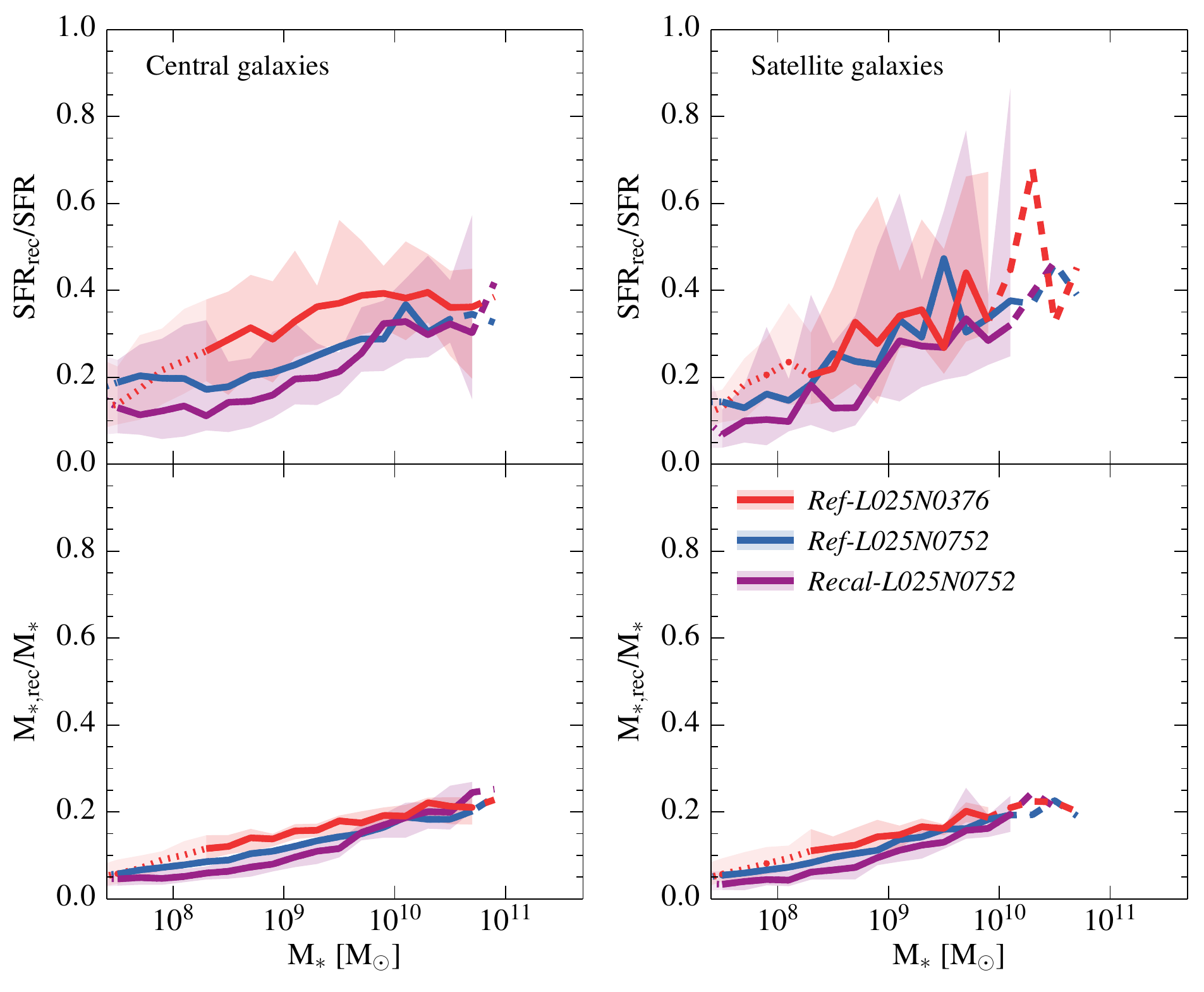}
\end{center}
\caption{Test for numerical convergence of the \sfrratio - $M\sub{\ast}$ (top) and \mstarratio - $M\sub{\ast}$ (bottom) relations (presented in Figs.~\ref{fig:massdep_eagle_reso} and \ref{fig:massdep_eagle}) for central (left) and satellite (right) galaxies at $z=0$, comparing an EAGLE simulation with the fiducial resolution (\emph{Ref-L025N376}; red) to two EAGLE simulations with $8$ times higher mass resolution, with (\emph{Recal-L025N0752}; purple) and without (\emph{Ref-L025N0752}; blue) recalibrated subgrid feedback parameters. All three simulations were run in volumes of size $L = 25$ cMpc. The curves and shaded regions have the same meaning as in Fig.~\ref{fig:massdep_eagle_reso}, but for clarity we only show the 10th to 90th percentile range for the \emph{Ref-L025N376} and \emph{Recal-L025N752} simulations. For galaxies with $M\sub{\ast} \gtrsim 10^{9.8}$ M$\sub{\odot}$, \sfrratio and \mstarratio as a function of stellar mass are reasonably well converged. For galaxies with $M\sub{\ast} \lesssim 10^{9.8}$ M$\sub{\odot}$, the fiducial EAGLE simulation likely overpredicts the SFR and stellar mass contributed by recycling, by at most a factor of $\sim 2$ ($\sim 0.3$ dex) at $M\sub{\ast} \sim 10^{9}$ M$\sub{\odot}$.}
\label{fig:reso_test}
\end{figure*}

We test for numerical convergence with respect to resolution of the \sfrratio - $M\sub{\ast}$ and \mstarratio - $M\sub{\ast}$ relations using a set of three EAGLE simulations that were run in volumes of size $L = 25$ cMpc. We consider both `weak' and `strong' convergence (following the nomenclature introduced by S15) by comparing simulations with recalibrated and non-recalibrated subgrid physics, respectively. We use:
\begin{itemize}
\item One simulation with $N = 376^3$ and with the same subgrid model parameters as our fiducial $L = 100$ cMpc, $N = 1504^3$ simulation that was used throughout this work. This simulation (denoted \emph{Ref-L025N376}) has the same resolution as the fiducial simulation.
\item One simulation with $N = 752^3$ and the same subgrid model parameters as the fiducial simulation, but with $8$ times higher mass resolution (\emph{Ref-L025N752}).
\item One simulation with $N = 752^3$ and with a \emph{recalibrated} set of subgrid model parameters for star formation feedback, AGN feedback and the accretion onto BHs, in order to improve the match with the observed $z \simeq 0$ GSMF at this $8$ times higher mass resolution (\emph{Recal-L025N752}). In short, the recalibration corresponds to a change in the density dependence of the stellar feedback efficiency parameter $f\sub{th}$, such that the feedback efficiency is increased in higher-density gas while keeping the average $f\sub{th}$ roughly equal to $1$. This is done in order to compensate for the increase in cooling losses, which arise as a result of the locally higher gas densities that are resolved in the higher-resolution model.
\end{itemize}
A comparison of these three EAGLE simulations is shown in Fig.~\ref{fig:reso_test}, where we plot \sfrratio (top) and \mstarratio (bottom) as a function of stellar mass. We show these relations for both central galaxies (left) and satellite galaxies (right). Since our conclusions about resolution convergence are broadly consistent between centrals and satellites (although with somewhat poorer sampling for the latter), we will focus the discussion below on central galaxies.

Comparing the fiducial resolution simulation (\emph{Ref-L025N0376}; red) to the two higher-resolution simulations, with (\emph{Recal-L025N0752}; purple) and without (\emph{Ref-L025N0752}; blue) recalibrated subgrid feedback parameters, we infer that for central galaxies with $M\sub{\ast} \lesssim 10^{9.8}$ M$\sub{\odot}$, \sfrratio and \mstarratio as a function of stellar mass are not numerically converged at the fiducial resolution. The `strong' convergence is somewhat better than the `weak' convergence. At $M\sub{\ast} \sim 10^{9}$ M$\sub{\odot}$, \sfrratio and \mstarratio in \emph{Ref-L025N0376} are almost a factor of $2$ ($0.3$ dex on a logarithmic scale) higher than in \emph{Recal-L025N0752}. This is not surprising considering the level of agreement between \emph{Ref-L100N1504} and \emph{Recal-L025N0752} for the mass-metallicity relations, where the latter is in better agreement with the observations (see fig 13. of S15).

At masses $M\sub{\ast} \gtrsim 10^{9.8}$ M$\sub{\odot}$, the relation between \mstarratio and stellar mass is fully converged (both `weakly' and `strongly'), while \sfrratio as a function of stellar mass shows substantial overlap between \emph{Ref-L025N0376}, \emph{Ref-L025N0752} and \emph{Recal-L025N0752}. Due to the small box size, however, \sfrratio is not well-sampled around $M\sub{\ast} \sim 10^{10.5}$ M$\sub{\odot}$, the mass scale at which \sfrratio reaches a maximum in our fiducial $L = 100$ cMpc model (see Fig.~\ref{fig:massdep_eagle_reso}).

\subsection{Relation between recycled gas contributions and metallicity}
\label{sec:resotests_metal}

\begin{figure*}
\begin{center}
\includegraphics[width=0.95\textwidth]{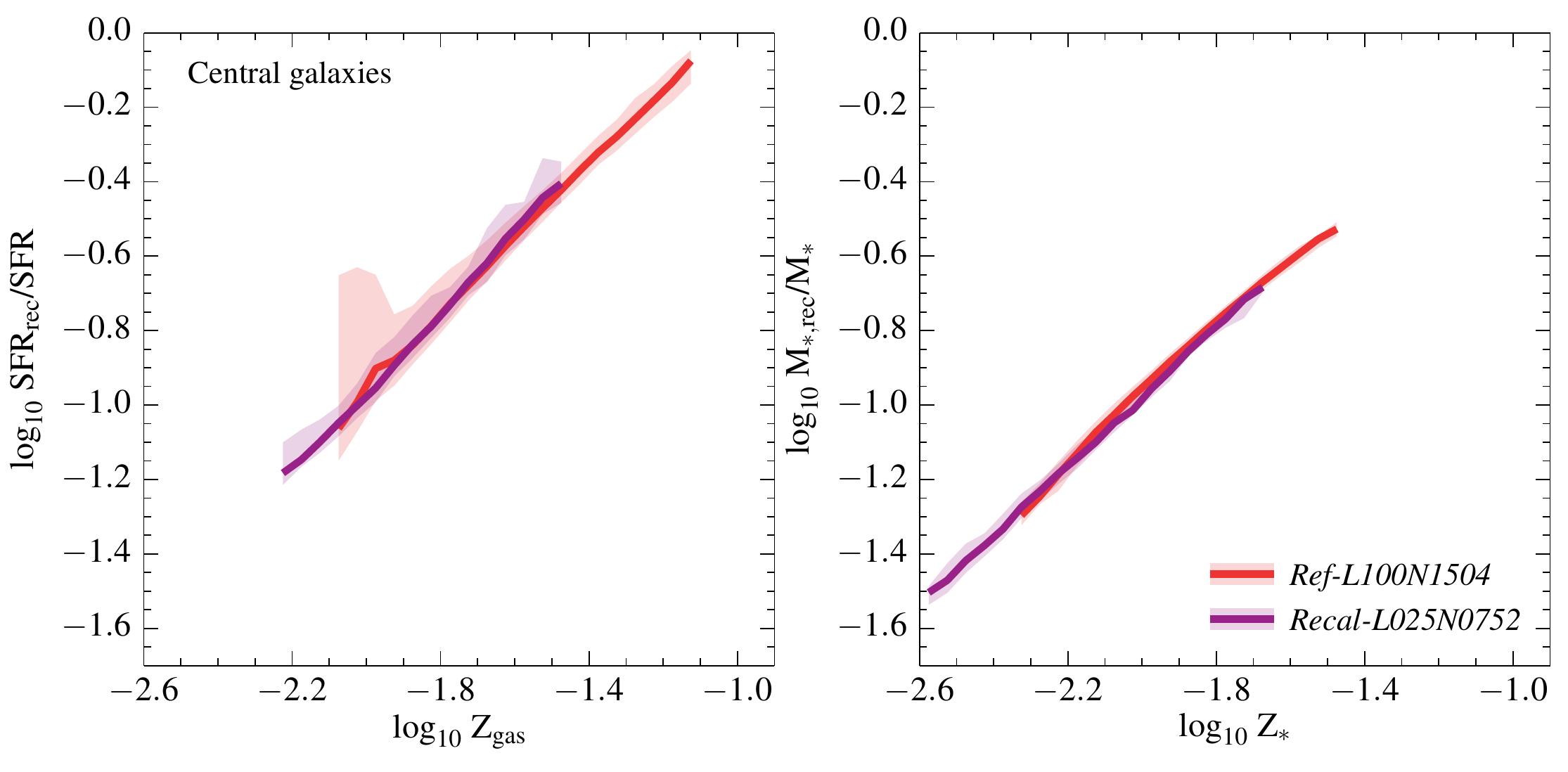}
\end{center}
\caption{Test for numerical convergence of the \sfrratio - $Z\sub{gas}$ (left) and \mstarratio - $Z\sub{\ast}$ (right) relations (presented in Fig.~\ref{fig:metallicity}) for central galaxies at $z=0$, comparing the fiducial EAGLE model (\emph{Ref-L100N1504}; red) and the high-resolution, recalibrated model (\emph{Recal-L025N0752}; purple). We only consider galaxies with stellar masses corresponding to at least $100$ gas particles, at the respective resolution. In the left panel we only consider subhaloes with a non-zero SFR. The curves show the median value in each logarithmic metallicity bin of size $0.05$ dex, if it contains at least $10$ galaxies. The shaded regions mark the $10$th to $90$th percentile ranges. The \sfrratio - $Z\sub{gas}$ and \mstarratio - $Z\sub{\ast}$ relations are converged at the fiducial resolution over the whole metallicity range.}
\label{fig:reso_test_metal}
\end{figure*}

Fig.~\ref{fig:reso_test_metal} shows the numerical convergence test of the \sfrratio - $Z\sub{gas}$ (left) and \mstarratio - $Z\sub{\ast}$ (bottom) relations for central galaxies, comparing the fiducial EAGLE model (\emph{Ref-L100N1504}; red) and the high-resolution, recalibrated model (\emph{Recal-L025N0752}; purple). While \emph{Recal-L025N0752} spans a metallicity range that is shifted towards somewhat lower values with respect to \emph{Ref-L100N1504}, due to its smaller box size and $8$ times higher mass resolution (we select stellar masses corresponding to at least $100$ gas particles at each resolution), the \sfrratio - $Z\sub{gas}$ and \mstarratio - $Z\sub{\ast}$ relations are converged with resolution over the whole metallicity range probed here. Where \emph{Recal-L025N0752} and \emph{Ref-L100N1504} overlap, their medians agree to better than $0.05$ dex in \sfrratio and to better than $0.04$ dex in \mstarratio.

One might wonder whether the secondary dependence on $\alpha$-enhancement (hence, implicitly on stellar mass; see Fig.~\ref{fig:ab_ratio}) affects the convergence of these relations. However, the dependence on stellar mass only becomes significant for $M\sub{\ast} \gtrsim 10^{10.5}$ M$\sub{\odot}$, which is the regime where \emph{Ref-L100N1504} and \emph{Recal-L025N0752} are converged (in terms of \mstarratio and $Z\sub{\ast}$) or at least broadly consistent (in terms of \sfrratio and $Z\sub{gas}$). As a consistency check, we repeat the calculation of \sfrratio and \mstarratio by applying the relations between recycling and metallicity to the observed mass-metallicity relations (as done in Section~\ref{sec:massdep_cen}) at higher resolution using \emph{Recal-L025N0752}. We find agreement with the results from \emph{Ref-L100N1504} to better than a factor of $\sim 1.07$ ($0.03$ dex) over the whole stellar mass range.


\section{Effect of using a 3D aperture}
\label{sec:aperture}

\begin{figure*}
\begin{center}
\includegraphics[width=0.8\textwidth]{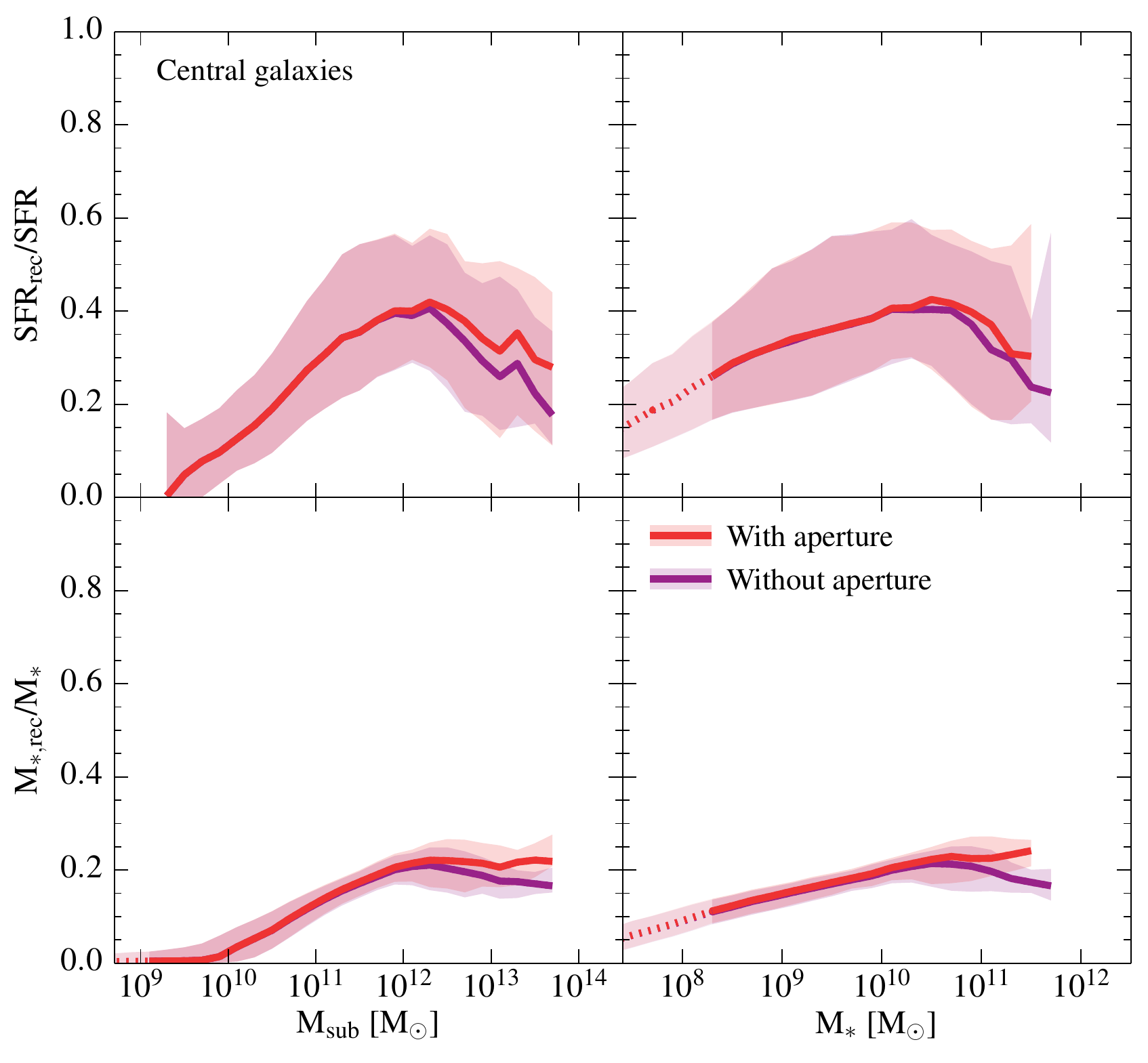}
\end{center}
\caption{The effect of using a $30$ pkpc 3D aperture, comparing results from the EAGLE fiducial model, \emph{Ref-L100N1504}, with (red) and without (purple) applying the aperture in calculating galaxy properties. The curves show the fraction of the SFR (top) and stellar mass (bottom) contributed by recycling for central galaxies at $z=0$ as a function of their subhalo mass (left) and stellar mass (right). The curves and shaded regions have the same meaning as in Fig.~\ref{fig:massdep_eagle_reso}. The effect of using an aperture is a change in the slope of the two recycled gas fractions at $M\sub{\ast} \gtrsim 10^{11}$ M$\sub{\odot}$ ($M\sub{sub} \gtrsim 10^{12.5}$ M$\sub{\odot}$), which becomes somewhat shallower. For the fraction of the stellar mass contributed by recycling this results in a roughly flat trend instead of a decrease with subhalo and stellar mass, hence mitigating the effect of increasing AGN feedback efficiency.}
\label{fig:aper_comp}
\end{figure*}

Fig.~\ref{fig:aper_comp} shows the effect of using a $30$ pkpc 3D aperture on the contribution of recycled gas to the SFR (top) and stellar mass (bottom) in central galaxies at $z=0$ as a function of their subhalo mass (left) and stellar mass (right). We compare results from the EAGLE fiducial model, \emph{Ref-L100N1504}, with (red) an without (purple) the aperture. Recall that the aperture only applies to galaxy properties, hence the subhalo mass is not affected.

While the majority of the star formation takes places within the central $30$ pkpc, causing the effect of the aperture on the total SFR to be small, there is still an enhancement in \sfrratio (upper panels) inside the aperture compared to its value over the whole galaxy. This enhancement is consistent with recycling-fuelled star formation being more important in the central parts of galaxies (see Fig.~\ref{fig:massdep_radius}). The effect is significant for $M\sub{\ast} \gtrsim 10^{11}$ M$\sub{\odot}$ ($M\sub{sub} \gtrsim 10^{12.5}$ M$\sub{\odot}$) and increases with mass up to a difference of a factor of $\sim 1.5$ ($\sim 0.18$ dex) in the left panel, whereas in the right panel the effect is smaller (up to a factor of $\sim 1.2$, or $\sim 0.08$ dex) due to the simultaneous decrease in the stellar mass if an aperture is applied.

Similarly, in the lower panels, \mstarratio is enhanced if an aperture is applied. In this case, the effect of using an aperture is that the decrease of \mstarratio with mass for $M\sub{\ast} \gtrsim 10^{10.5}$ M$\sub{\odot}$ ($M\sub{sub} \gtrsim 10^{12.2}$ M$\sub{\odot}$) becomes a flattening at a roughly constant value ($\sim 22 \%$ instead of $17 \%$ at $M\sub{\ast} \sim 10^{11.5}$ M$\sub{\odot}$).

\bsp
\label{lastpage}
\end{document}